# Vacuum Technology for Ion Sources

*P. Chiggiato*
CERN, Geneva, Switzerland

**Abstract**
The basic notions of vacuum technology for ion sources are presented, with emphasis on pressure profile calculation and choice of pumping technique. A Monte Carlo code (Molflow+) for the evaluation of conductances and the vacuum–electrical analogy for the calculation of time-dependent pressure variations are introduced. The specific case of the Linac4 H⁻ source is reviewed.

## 1 Introduction

Ion sources are among the very few places in accelerators where gas is intentionally injected during operation. Therefore, the first goal of the vacuum system is to remove the injected gas before it reaches the rest of the acceleration chain, where a lower gas density is required. The injected and then evacuated gas flow is in general very high. Such a requirement, together with the long uninterrupted operational time, imposes a drastic constraint on the selection of the vacuum pumps and the way they are operated.

Mechanical pumps are the obvious choice close to the injection points, where the gas throughput is the highest. Such pumps evacuate the gas molecules outside the ion source. However, as soon as most of the injected gas is evacuated and lower pressures must be reached, the vacuum system should rely on capture pumps because they are in general more reliable than mechanical pumps.

The evaluation of the dynamic and static pressure profiles should be undertaken in parallel with the mechanical design. Basic calculations may be carried out for simple geometries. However, the complexity of the mechanical design requires advanced computation. The molecular trajectories in the vacuum system are simulated by Monte Carlo codes, and pressures are evaluated by counting the molecular collisions with virtual surfaces. The analogy between vacuum systems and electrical networks, in conjunction with the Monte Carlo simulations, may be used to obtain time-dependent pressure profiles.

In ion sources, pressure is measured by applying different principles. In the highest pressure range, near the injection points, mechanical and thermal conductivity effects are still significant in the gas phase. In this case, thermal and capacitance pressure gauges are employed. For a more rarefied gas, pressure readings are obtained by measuring gas-ion currents intentionally produced by electron impact ionization.

This chapter focuses mainly on high-flow gas pumping and the calculation of pressure profiles. The first pages are dedicated to the presentation of the basic knowledge and terminology of vacuum technology, especially conductance and pumping speed. Then the main pumping mechanisms are explained, with the focus on operational details that are relevant to ion sources. Pressure measurement is not included because, in general, it is not considered as a critical issue in ion sources. A comprehensive introduction to vacuum technology may be found in the books listed in the references.

## 2    Basic notions of vacuum technology

In the framework of vacuum technology for particle accelerators, a rarefied gas in equilibrium is always described by the ideal-gas equation of state:

$$PV = N_{\text{moles}}RT, \qquad (1)$$

where $P$, $T$ and $V$ are the gas pressure, temperature and volume, respectively, and $R$ the ideal-gas constant (8.314 J K$^{-1}$ mol$^{-1}$ in SI units). From statistical physics considerations, Eq. (1) may be rewritten in terms of the total number of molecules $N$ in the gas:

$$PV = Nk_{\text{B}}T, \qquad (2)$$

where $k_{\text{B}}$ is the Boltzmann constant ($1.38 \times 10^{-23}$ J K$^{-1}$ in SI units).

In the International System of Units (SI), the pressure is reported in pascals (Pa): 1 Pa is equivalent to the pressure exerted by one newton on a square metre (1 N m$^{-2}$). Other units are regularly used in vacuum technology, in particular bar and its submultiple the millibar (mbar). The Torr is still occasionally used, mostly in the USA; it is equivalent to the pressure exerted by a 1 mm high column of Hg. The conversion values between the common pressure units are shown in Table 1.

**Table 1:** Conversion values for the most common pressure units of vacuum technology.

|         | Pa                  | bar                  | atm                   | Torr                |
|---------|---------------------|----------------------|-----------------------|---------------------|
| 1 Pa    | 1                   | $10^{-5}$            | $9.87 \times 10^{-6}$ | $7.5 \times 10^{-3}$ |
| 1 bar   | $10^{5}$            | 1                    | 0.987                 | 750.06              |
| 1 mbar  | $10^{2}$            | $10^{-3}$            | $0.967 \times 10^{-3}$ | 0.75                |
| 1 atm   | $1.013 \times 10^{5}$ | 1.013              | 1                     | 760                 |
| 1 Torr  | 133.32              | $1.33 \times 10^{-3}$ | $1.32 \times 10^{-3}$ | 1                   |

The number density ($n = N/V$) of gas molecules is easily calculated by Eq. (2). Some typical examples are collected in Table 2.

**Table 2:** Typical number density at room temperature and helium boiling point.

|                                                      | Pressure (Pa)         | 293 K (molecules cm$^{-3}$) | 4.3 K (molecules cm$^{-3}$) |
|------------------------------------------------------|-----------------------|-----------------------------|-----------------------------|
| Atmospheric pressure at sea level                    | $1.013 \times 10^{5}$ | $2.5 \times 10^{19}$        | $1.7 \times 10^{21}$        |
| Typical plasma chambers                              | 1                     | $2.5 \times 10^{14}$        | $1.7 \times 10^{16}$        |
| Linac pressure upper limit                           | $10^{-5}$             | $2.5 \times 10^{9}$         | $1.7 \times 10^{11}$        |
| Lowest pressure ever measured at room temperature [1] | $10^{-12}$            | 250                         | $1.7 \times 10^{4}$         |

In addition to the number of molecules or moles, gas quantities are expressed as pressure–volume (*PV*) values at a given temperature. Pressure–volume quantities are converted to number of molecules by dividing them by $k_B T$ as given in the equation of state. For example, 1 Pa m³ at 293 K is equivalent to

$$N = \frac{1 \, [\text{Pa m}^3]}{1.38 \times 10^{-23} \, [\text{J K}^{-1}] \times 293 \, [\text{K}]} = 2.47 \times 10^{20} \text{ molecules.}$$

In the same way, at 293 K, 1 Torr litre = $3.3 \times 10^{19}$ molecules and 1 mbar litre = $2.47 \times 10^{19}$ molecules. In vacuum technology, pressure–volume units are used most of the time to report gas quantities and flow rates. For example, the gas flow rate is generally reported in mbar litre s$^{-1}$ or Torr litre s$^{-1}$.

In vacuum systems, pressures span several orders of magnitude (Table 3). Degrees of vacuum are defined by upper and lower pressure boundaries. Different degrees of vacuum are characterized by different pumping technologies, pressure measurement, materials and surface treatments. Ion sources operate in the degrees of vacuum that are usually called medium and high vacuum.

**Table 3:** Degrees of vacuum and their pressure boundaries.

|  | Pressure boundaries (mbar) | Pressure boundaries (Pa) |
|---|---|---|
| Low vacuum (LV) | 1000–1 | $10^5$–$10^2$ |
| Medium vacuum (MV) | 1–$10^{-3}$ | $10^2$–$10^{-1}$ |
| High vacuum (HV) | $10^{-3}$–$10^{-9}$ | $10^{-1}$–$10^{-7}$ |
| Ultra-high vacuum (UHV) | $10^{-9}$–$10^{-12}$ | $10^{-7}$–$10^{-10}$ |
| Extreme vacuum (XHV) | <$10^{-12}$ | <$10^{-10}$ |

## 2.1 Gas kinetics

The kinetics of ideal-gas molecules is described by the Maxwell–Boltzmann theory [2]. For an isotropic gas, the model provides the probabilistic distribution of the molecular speed magnitudes. The mean speed of molecules $\langle v \rangle$, i.e. the mathematical average of the speed distribution, is given by

$$\langle v \rangle = \sqrt{\frac{8 k_B T}{\pi m}} = \sqrt{\frac{8 R T}{\pi M}}, \tag{3}$$

where *m* is the mass of the molecule and *M* is the molar mass. The unit of both masses is kg in SI. Typical mean speed values are shown in Table 4.

**Table 4:** Mean speed of gas molecules of different mass at room temperature and helium boiling point.

|  | $H_2$ | He | $CH_4$ | $N_2$ | Ar |
|---|---|---|---|---|---|
| $\langle v \rangle$ at 293 K (m s$^{-1}$) | 1761 | 1244 | 622 | 470 | 394 |
| $\langle v \rangle$ at 4.3 K (m s$^{-1}$) | 213 | 151 | 75 | 57 | 48 |

Another important result of the Maxwell–Boltzmann theory is the calculation of the molecular impingement rate $\varphi$ on a surface, i.e. the rate at which gas molecules collide with a unit surface area exposed to the gas. Assuming that the density of molecules all over the volume is uniform, it can be shown [3] that

$$\varphi = \frac{1}{4}n\langle v \rangle, \tag{4}$$

and using Eq. (3) for the mean speed as obtained by the Maxwell–Boltzmann theory,

$$\varphi = \frac{1}{4}n\sqrt{\frac{8k_\mathrm{B}T}{\pi m}}. \tag{5}$$

Numerical values in terms of $P$, $T$ and molar mass are obtained by Eq. (6); some selected values are reported in Table 5:

$$\varphi\,[\mathrm{cm}^{-2}\,\mathrm{s}^{-1}] = 2.635 \times 10^{22}\,\frac{P\,[\mathrm{mbar}]}{\sqrt{M\,[\mathrm{g}]\,T\,[\mathrm{K}]}}. \tag{6}$$

**Table 5:** Molecular impingement rates at room temperature for $H_2$, $N_2$ and Ar at some selected pressures.

| Gas | Pressure (mbar) | Impingement rate (cm$^{-2}$ s$^{-1}$) |
|---|---|---|
| $H_2$ | $10^{-3}$ | $1.1 \times 10^{18}$ |
|  | $10^{-8}$ | $1.1 \times 10^{13}$ |
|  | $10^{-14}$ | $1.1 \times 10^{7}$ |
| $N_2$ | $10^{-3}$ | $2.9 \times 10^{17}$ |
|  | $10^{-8}$ | $2.9 \times 10^{12}$ |
| Ar | $10^{-3}$ | $2.4 \times 10^{17}$ |
|  | $10^{-8}$ | $2.4 \times 10^{12}$ |

## 2.2  Mean free path and Knudsen number

In any physically limited vacuum system, molecules collide between each other and with the walls of the vacuum envelope/container. In the first case, the average length of the molecular path between two points of consecutive collisions, i.e. the mean free path $\bar{\lambda}$, is inversely proportional to the number density $n = P/k_\mathrm{B}T$ and the collision cross-section $\sigma_\mathrm{c}$ [4], as given by

$$\bar{\lambda} = \frac{1}{\sqrt{2}n\sigma_\mathrm{c}}. \tag{7}$$

For elastic collisions between hard spheres, Eq. (7) can be written in terms of the molecular diameter $\delta$ as

$$\bar{\lambda} = \frac{1}{\sqrt{2}\pi n\delta^2} = \frac{k_\mathrm{B}T}{\sqrt{2}\pi P\delta^2}. \tag{8}$$

The collision cross-sections for common gas species in vacuum systems are listed in Table 6.

**Table 6:** Elastic collision cross-sections for five different molecules.

| Gas | $H_2$ | He | $N_2$ | $O_2$ | $CO_2$ |
|---|---|---|---|---|---|
| $\sigma_c$ (nm²) | 0.27 | 0.27 | 0.43 | 0.40 | 0.52 |

For numerical purposes, Eq. (8) can be rewritten for a specific gas as a function of temperature and pressure. For example, for $H_2$, we obtain

$$\bar{\lambda}_{H_2}[m] = 4.3 \times 10^{-5} \frac{T[K]}{P[Pa]} \quad (9)$$

The values of the mean free path for $H_2$ at room temperature are shown in Fig. 1 as a function of the gas pressure.

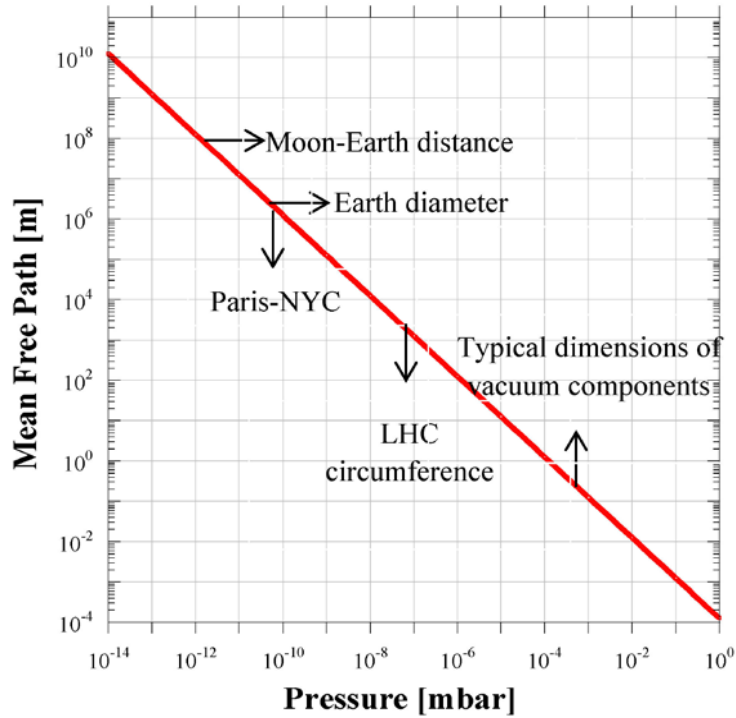

**Fig. 1:** Mean free path of $H_2$ molecules at 293 K

When the mean free path is of the order of typical dimensions of the vacuum vessel, for example, the diameter of cylindrical beam pipes, molecular collisions with the wall of the vacuum envelope become preponderant. For even longer $\bar{\lambda}$, the gas dynamics is dominated by molecule–wall collisions; intermolecular interactions lose any effect on the gas displacement.

The adimensional Knudsen number $K_n$ translates into numerical values the considerations expressed above. It is defined as the ratio between the mean free path and a characteristic dimension of a vacuum system (*D*):

$$K_n = \frac{\bar{\lambda}}{D} \quad (10)$$

The values of $K_n$ delimit three gas dynamic regimes as reported in Table 7.

**Table 7:** Gas dynamic regimes defined by the Knudsen number.

| $K_n$ range | Regime | Description |
| --- | --- | --- |
| $K_n > 0.5$ | Free molecular flow | Molecule–wall collisions dominate |
| $K_n < 0.01$ | Continuous (viscous) flow | Gas dynamics dominated by intermolecular collisions |
| $0.5 < K_n < 0.01$ | Transitional flow | Transition between molecular and viscous flow |

Typical beam pipe diameters are of the order of 10 cm. Therefore, the free molecular regime is obtained for pressures in the low $10^{-3}$ mbar range or lower. Except for the ion source's plasma chambers, the vacuum systems of accelerators operate in the free molecular regime; only this vacuum regime is considered in this chapter. Excellent introductions to vacuum systems in viscous and transitional regimes can be found in Refs. [5] and [6].

The free molecular flow regime characterizes and determines the pumping and pressure reading mechanisms that can be used in particle accelerators. Pumps and instruments must act on single molecules since there is no interaction between molecules. Collective phenomena such as pressure waves and suction do not influence gas dynamics in free molecular flow.

## 2.3 Conductance in free molecular flow

In the free molecular regime, the net gas flow $Q$ between two points of a vacuum system is proportional to the pressure difference $(P_1 - P_2)$ at the same points:

$$Q = C(P_1 - P_2), \tag{11}$$

where $C$ is called the gas conductance of the vacuum system between the two points. In the free molecular regime, the conductance does not depend on pressure. It depends only on the mean molecular speed and the vacuum system geometry. If the gas flow units are expressed in terms of pressure–volume (e.g. mbar litre s$^{-1}$ or Pa m$^3$ s$^{-1}$), the conductance is reported as volume per unit time (i.e. l s$^{-1}$ or m$^3$ s$^{-1}$, from here onward litre are reported with the letter l).

The conductance is easily calculated for the simplest geometry, i.e. a small wall slot of surface area $A$ and infinitesimal thickness dividing two volumes of the same vacuum system (see Fig. 2).

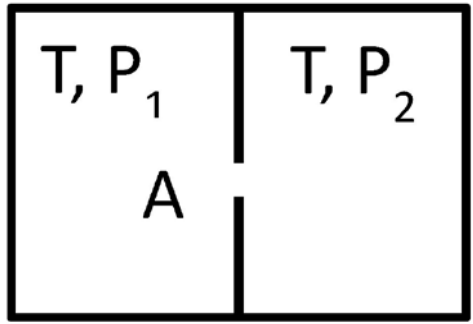

**Fig. 2:** Schematic drawing of two volumes communicating through a thin and small wall slot

The net flow of molecules from one volume to the other may be calculated by the molecular impingement rate given by Eq. (4). The number of molecules of volume 1 that go into volume 2 ($\varphi_{1 \to 2}$) is

$$\varphi_{1 \to 2} = \frac{1}{4} A n_1 \langle v \rangle$$

while that from volume 2 to volume 1 is

$$\varphi_{2\to1} = \frac{1}{4} A n_2 \langle v \rangle.$$

The net molecular flow is given by the difference between the two contributions,

$$\varphi_{1\to2} - \varphi_{2\to1} = \frac{1}{4} A (n_1 - n_2) \langle v \rangle,$$

and from Eqs. (2) and (3)

$$\varphi_{1\to2} - \varphi_{2\to1} = \frac{1}{4} A \frac{\langle v \rangle}{k_B T} (P_1 - P_2).$$

Multiplying both terms of the equality by $k_B T$ and applying Eq. (2), the gas flow in pressure–volume units is obtained as

$$Q = \frac{1}{4} A \langle v \rangle (P_1 - P_2). \tag{12}$$

Comparing Eqs. (11) and (12), it comes out that the conductance of the wall slot is proportional to the surface area of the slot and the mean speed of the molecules:

$$C = \frac{1}{4} A \langle v \rangle \propto \sqrt{\frac{T}{m}}. \tag{13}$$

From Eqs. (3) and (13), it can be shown that the conductance of the wall slots is inversely proportional to the square root of the molecular mass. Therefore, for equal pressure drop, the gas flow of $H_2$ is the highest. Finally, for gas molecules of different masses, the conductance scales as the square root of the inverse mass ratio:

$$\frac{C_1}{C_2} = \sqrt{\frac{m_2}{m_1}}. \tag{14}$$

As an example, the conductance for $N_2$ is $\sqrt{2/28} = 0.27$ times that for $H_2$, namely 3.7 times lower. Table 8 collects conductance values, for an orifice, per unit surface area ($C'$) at room temperature for common gas species.

**Table 8:** Unit surface area conductances for common gas species in two different units.

| Gas | $H_2$ | He | $CH_4$ | $H_2O$ | $N_2$ | Ar |
|---|---|---|---|---|---|---|
| $C'$ at 293 K (m³ s⁻¹ m⁻²) | 440.25 | 311 | 155.5 | 146.7 | 117.5 | 98.5 |
| $C'$ at 293 K (l s⁻¹ cm⁻²) | 44 | 31.1 | 15.5 | 14.7 | 11.75 | 9.85 |

For more complex geometries than wall slots, the transmission probability $\tau$ is introduced. If two vessels, at the same temperature, are connected by a duct (see Fig. 3 for symbols), the gas flow from vessel 1 to vessel 2 ($\varphi_{1\to2}$) is calculated multiplying the number of molecules impinging on the entrance section of the duct by the probability $\tau_{1\to2}$ for a molecule to be transmitted into vessel 2 without coming back to vessel 1:

$$\varphi_{1\to2} = \frac{1}{4} A_1 n_1 \langle v \rangle \, \tau_{1\to2}. \tag{15}$$

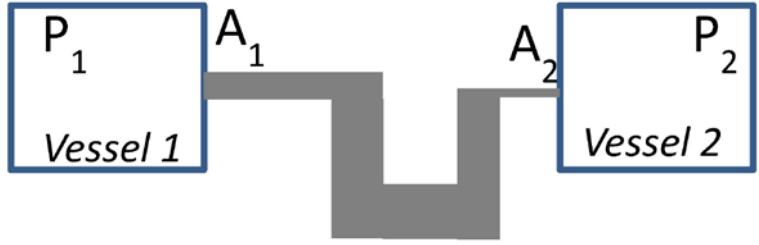

**Fig. 3:** Schematic drawing of two vessels connected by a duct

Similarly, the gas flow from vessel 2 to vessel 1 is written as

$$\varphi_{2\to1} = \frac{1}{4} A_2 n_2 \langle v \rangle \, \tau_{2\to1}. \tag{16}$$

In the absence of net flow, $\varphi_{1\to2} = \varphi_{2\to1}$ and $n_1 = n_2$, then

$$A_1 \tau_{1\to2} = A_2 \tau_{2\to1}. \tag{17}$$

When $n_1 \ne n_2$, a net flow is set up. It can be calculated by taking into account Eqs. (15)–(17) and (2):

$$\varphi_{1\to2} - \varphi_{2\to1} = \frac{1}{4} A_1 \langle v \rangle \tau_{1\to2} \frac{(P_1 - P_2)}{k_\mathrm{B} T}. \tag{18}$$

In pressure–volume units, Eq. (18) becomes

$$Q = \frac{1}{4} A_1 \langle v \rangle \tau_{1\to2} (P_1 - P_2) = C' A_1 \tau_{1\to2} (P_1 - P_2), \tag{19}$$

where, as already mentioned, $C'$ is the conductance of the unit surface area wall slot. Comparing Eqs. (11) and (19), it can be seen that the conductance of the connecting duct is equal to the conductance of the duct entrance in vessel 1, considered as a wall slot, multiplied by the molecular transmission probability from vessel 1 to vessel 2:

$$C = C' A_1 \tau_{1\to2}. \tag{20}$$

### 2.3.1 *Evaluation of the transmission probability*

The transmission probabilities depend only on vacuum component geometry. They may be calculated analytically for simple geometry by means of relatively complex integral equations (Clausing equations, see Ref. [7]). Approximate formulas are reported in many vacuum-technology books, for example in Ref. [8].

For the very common case of tubes of uniform circular cross-section of length $L$ and radius $R$, the Santeler equation [9] gives the transmission probability with less than 0.7% error:

$$\tau = \tau_{1\to2} = \tau_{2\to1} = \frac{1}{1 + \frac{3L}{8R}\left(1 + \frac{1}{3\left(1 + \frac{L}{7R}\right)}\right)}. \tag{21}$$

For long tubes, i.e. $L/R \gg 1$, Eq. (21) can be simplified to

$$\tau \approx \frac{1}{1 + \frac{3L}{8R}} \approx \frac{8}{3}\frac{R}{L}. \tag{22}$$

Equation (22) combined with Eq. (20) gives the conductance of long circular pipes; it is one of the most used equations in vacuum technology. For $N_2$ it may be written as

$$C \approx 11.75 \times \frac{\pi D^2}{4} \times \frac{4D}{3L} = 12.3 \frac{D^3}{L} \; [\text{l s}^{-1}] \; ([D] \text{ and } [L] = \text{cm}). \tag{23}$$

As a result, the conductance of a tube is strongly dependent on its diameter.

The Santeler equation and its approximation for long tubes are plotted in Fig. 4. The latter can be applied with errors less than 10% for $L/R \gg 20$. As shown in the figure, the transmission probability is about 0.5 for circular tubes for which the diameter is equal to their length. In other words, the conductance of such tubes is half that of their entrance surface.

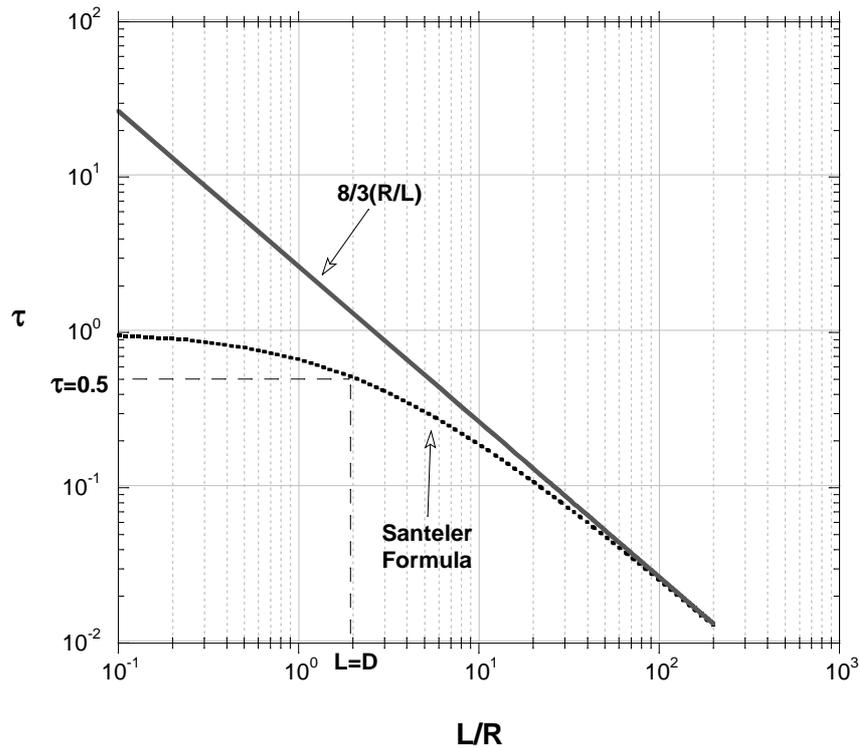

**Fig. 4**: Transmission probability of tubes of uniform circular cross-section calculated by the Santeler equation and its approximation for high *L/R*.

Conductances of more complicated components are calculated by test-particle Monte Carlo methods (TPMC). The components are modelled in three dimensions; the TPMC codes generate molecules at the entrance of the component pointing in 'random' directions according to the cosine distribution. When the molecules impinge on the internal wall of the component, they are re-emitted again randomly. The program follows the molecular traces until they reach the exit of the component. The transmission probability is given by the ratio between the number of 'escaped' and 'injected' molecules [10]. Many simulated molecules are needed to reduce the statistical scattering.

The reference TPMC software at CERN is MolFlow+ [11]. This powerful tool for high-vacuum applications imports the 3D drawing of the vacuum components and generates 'random' molecules on any surface of interest. Figure 5 shows the model and the molecular tracks for the extraction chamber of the Linac4's ion source.

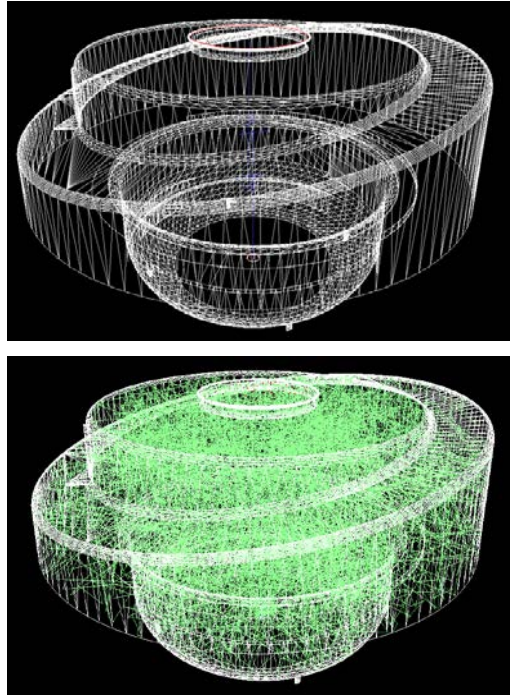

**Fig. 5**: Meshes used for the Monte Carlo simulation of the extraction chamber of Linac4. The second picture shows molecular tracks (in green) generated by the MolFlow+ code.

### *2.3.2 Combination of conductances*

Elementary vacuum components are installed either in series, i.e. traversed by the same net gas flow, or in parallel, i.e. equal pressures at the extremities (Fig. 6).

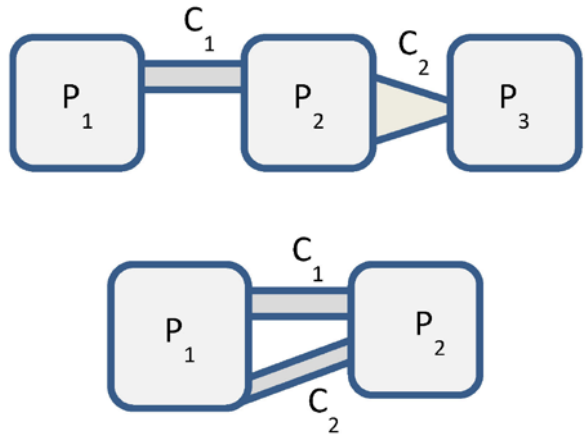

**Fig. 6:** Schematic drawings of components installed in series (top) and in parallel (bottom)

With reference to Fig. 6, the net gas flow in the two components connected *in series* is given by

$$Q = C_1(P_1 - P_2),$$
$$Q = C_2(P_2 - P_3). \qquad (24)$$

A total conductance $C_{\text{TOT}}$ equivalent to $C_1$ and $C_2$ is introduced in such a way that

$$Q = C_{\text{TOT}}(P_1 - P_3). \qquad (25)$$

Combining Eqs. (24) and (25), $C_{\text{TOT}}$ is easily calculated as

$$\frac{1}{C_{\text{TOT}}} = \frac{1}{C_1} + \frac{1}{C_2}. \tag{26}$$

In general, for $N$ components in series:

$$\frac{1}{C_{\text{TOT}}} = \sum_1^N \frac{1}{C_i}. \tag{27}$$

In the same way, it can be shown that for $N$ components installed *in parallel*, the total conductance is the sum of the conductances of all components:

$$C_{\text{TOT}} = \sum_1^N C_i. \tag{28}$$

### 2.4 Pumping speed

In vacuum technology, a pump is any 'object' that removes gas molecules from the gas phase. A vacuum pump is characterized by its pumping speed $S$, which is defined as the ratio between the pumped gas flow $Q_P$ (pump throughput) and the pump inlet pressure $P$:

$$S = \frac{Q_P}{P}. \tag{29}$$

The pumping speed unit is volume over time, thus the same unit as conductance. In high vacuum, $S$ is in general expressed in $l\ s^{-1}$ or $m^3\ s^{-1}$; in low and medium vacuum, $m^3\ min^{-1}$ is used. In a more general way, $S$ can be defined as the derivative of the pump throughput with respect to the pump inlet pressure:

$$S = \frac{\partial Q_P}{\partial P}. \tag{30}$$

The pump throughput can be written as the gas flow through the cross-section of the pump inlet (surface area $A_P$) multiplied by the capture probability $\sigma$, i.e. the probability for a molecule that enters the pump to be definitely removed and never more to reappear in the gas phase of the vacuum system (see Eq. (31)). In the literature, $\sigma$ is also called the Ho coefficient:

$$Q_P = \varphi A_P \sigma = \frac{1}{4} A_P n \langle v \rangle \sigma. \tag{31}$$

Considering Eqs. (13) and (2), it turns out that

$$Q_P = A_P C' n \sigma = A_P C' \sigma \frac{P}{k_B T}.$$

From the definition of the pumping speed and converting the throughput into pressure–volume units:

$$S = A_P C' \sigma. \tag{32}$$

Therefore, the pumping speed is equal to the conductance of the pump inlet cross-section multiplied by the capture probability. The maximum theoretical pumping speed of any pump is obtained for $\sigma = 1$ and it is equal to the conductance of the pump inlet cross-section. Table 9 reports some values of the maximum pumping speed of lump pumps for typical diameters of the pump inlet. Because $S$ depends on $C'$, and so on the inverse of the square root of the molecular mass, the maximum theoretical pumping speed is that for $H_2$.

The pumping speed given by the suppliers is called the *nominal pumping speed*; it refers to the pump inlet. The *effective pumping speed* $S_{eff}$ is that acting directly in the vacuum vessel of interest. The effective pumping speed is lower than the nominal pumping speed owing to gas flow restrictions interposed between the pump and the vessel.

**Table 9:** Maximum pumping speeds in l s$^{-1}$ for different diameters of circular pump inlets.

| ID (mm) | $H_2$ | $N_2$ | Ar |
|---|---|---|---|
| 36 | 448 | 120 | 100 |
| 63 | 1371 | 367 | 307 |
| 100 | 3456 | 924 | 773 |
| 150 | 7775 | 2079 | 1739 |

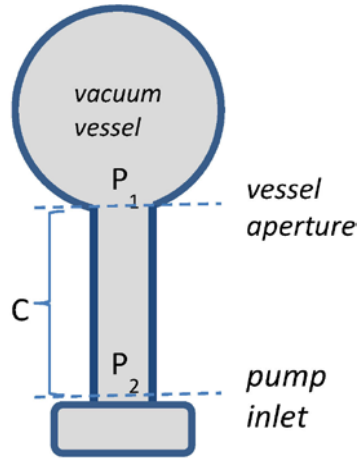

**Fig. 7**: Schematic drawing of a gas flow restriction of conductance $C$ interposed between a pump of pumping speed $S$ and a vacuum vessel.

The effective pumping speed is calculated considering the gas flow from the vessel and the pump. Taking into account Eqs. (11) and (29), with reference to Fig. 7, one obtains

$$Q = C_1(P_1 - P_2) = SP_2 = S_{eff}P_1$$

and so

$$\frac{1}{S_{eff}} = \frac{1}{S} + \frac{1}{C}. \tag{33}$$

As a result, for $C \ll S$, one finds $S_\text{eff} \approx C$. In other words, the effective pumping speed does not depend on the installed pump if the conductance of the interposed connection is very low. This conclusion is of primary importance in the design of efficient vacuum systems.

## 2.5  Outgassing

Several gas sources contribute to the total gas load in a vacuum system. In addition to intentional gas injections and air leaks, outgassing of materials plays a crucial role in the global gas balance. Materials release gas molecules that are both adsorbed onto their surfaces and dissolved in their bulk [12]. The outgassing is distributed uniformly all over the vacuum exposed surfaces; its rate is in general reported for unit surface area ($q$).

The outgassing rate is reduced by dedicated surface treatments. For metals, in general, several surface treatments are available, each aiming at removing a specific set of contaminants. Gross contamination and the sorption layer (hydrocarbons, Cl compounds, silicone greases, etc.) are removed by solvent and detergent cleaning. Solvent molecules interact with contaminants and transport them away from the surface by diffusion. They are in general quite selective: perchloroethylene ($C_2Cl_4$) has a wide spectrum of applications, while hydrofluorocarbon (HFC) has a more restricted action. Detergent molecules are dissolved in water; they are based on an aggregate of surfactant (surface-acting agent) molecules, called a micelle, with hydrophilic and lipophilic extremities; they allow organic molecules and water to combine. Detergents are less selective than solvents. Thick oxide layers (1–10 nm thick) are removed by chemical pickling. The damaged skin of metals – full of dislocations, voids and impurities – is removed by acid etching or electropolishing.

For all materials, after thorough surface treatment, water vapour dominates the outgassing process. Its outgassing rate is inversely proportional to the pumping time ($t$) for smooth metals. The following empirical relationship is in general applied:

$$q_{H_2O} \approx \frac{3 \times 10^{-9}}{t\,[\text{h}]} \left[\frac{\text{mbar l}}{\text{s cm}^2}\right]. \tag{34}$$

For organic materials, in particular polymers, in the first phase of pumping, the water vapour outgassing rate is inversely proportional to the square root of the pumping time $q_{H_2O} \propto 1/\sqrt{t}$, the absolute value being strongly dependent on the nature of the material. The square root dependence indicates that water molecule diffusion in the bulk is the leading process.

The water vapour outgassing is reduced by *in situ* heating in vacuum (bake-out). Such a thermal treatment is very effective for metals if it is carried out for at least 12 h at temperatures higher than 120°C. After bake-out, $H_2$ becomes the main outgassed species for metals; the outgassing rate can be regarded as constant at room temperature. It can be reduced by heat treatment at higher temperature either *in situ* or *ex situ* in a vacuum furnace [12].

As shown in Table 10, the value of outgassing rate spans several orders of magnitude. The right choice of materials and treatments is essential in the design of vacuum systems to limit the gas charge. The outgassing features should be taken into account at the very beginning of the design process when materials are selected.

**Table 10:** Values of outgassing rates for selected materials used in vacuum technology. The reported heat treatments are carried out both *in situ* and in vacuum.

| Material | q (mbar l s$^{-1}$ cm$^{-2}$) | Main gas species |
|---|---|---|
| Neoprene, not baked, after 10 h of pumping [13] | order of $10^{-5}$ | $H_2O$ |
| Viton, not baked, after 10 h of pumping [13] | order of $10^{-7}$ | $H_2O$ |
| Austenitic stainless steel, not baked, after 10 h of pumping | $3 \times 10^{-10}$ | $H_2O$ |
| Austenitic stainless steel, baked at 150°C for 24 h | $3 \times 10^{-12}$ | $H_2$ |
| OFS copper, baked at 200°C for 24 h | order of $10^{-14}$ | $H_2$ |

## 3 Calculation of pressure profiles

The calculation of the pressure profile along vacuum systems is an essential task of vacuum experts and should be tackled at the design phase. In general, the contributions to the total pressure of localized and distributed gas sources are considered separately and finally added. This is possible because in most cases the equations that describe pressure profiles are linear. This may not be true if the pumping speed is pressure-dependent.

### 3.1 Pressure profiles generated by localized gas sources

The pressure in a vacuum vessel is obtained by taking into account Eq. (29) and the intrinsic pressure limitation $P_0$ of the installed pumping system:

$$P = \frac{Q}{S} + P_0. \tag{35}$$

The pumping speed $S$ is either given by the supplier or preliminarily measured; $P_0$ is the pressure attained in the system without any gas load. When a restriction of conductance $C$ is interposed between the pump and the vessel, the effective pumping speed $S_{\text{eff}}$ is considered and Eq. (35) becomes

$$P = \frac{Q}{S_{\text{eff}}} + P_0 = \frac{Q(C + S)}{CS} + P_0. \tag{36}$$

When many vessels are interconnected, the flow balance is written in each vessel (node analysis). This analysis leads to a system of linear equations, from which the pressure values in each vessel are calculated. As an example, with reference to Fig. 8, in the first vessel, the injected gas flow ($Q$) is either pumped ($P_1 S_1$) or transmitted to the second vessel ($C_1(P_1 - P_2)$). This latter flow is pumped in the second vessel or transmitted to the third vessel, and so on. Thus

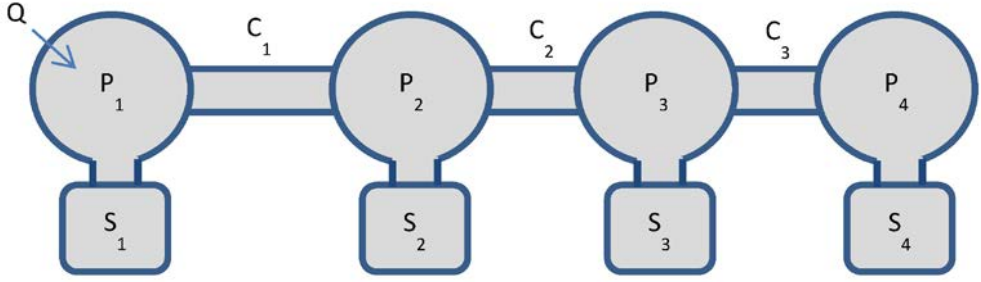

**Fig. 8**: Schematic drawing of four interconnected vacuum vessels. In each vessel, the gas flow balance is written (node analysis).

$$\begin{aligned} Q &= P_1 S_1 + C_1(P_1 - P_2), \\ C_1(P_1 - P_2) &= C_2(P_2 - P_3) + P_2 S_2, \\ C_2(P_2 - P_3) &= C_3(P_3 - P_4) + P_3 S_3, \\ C_3(P_3 - P_4) &= P_4 S_4. \end{aligned} \quad (37)$$

When a second localized gas flow is settled, the node analysis is repeated. The contributions of each localized gas flow to the pressure values are then added. If the cross-section of the interconnection ducts is constant, it can be shown that the pressure varies linearly between two connected vessels.

### 3.2 Pressure profiles generated by distributed gas sources

The pressure profiles generated by distributed outgassing and lump pumps is calculated analytically for simple geometries. Let us consider the case of a cylindrical beam pipe (radius $R$, length $L$) connected to a pump at one of its extremities. The calculation starts by imposing a mass balance equation in a generic small fraction of the beam pipe (see Fig. 9):

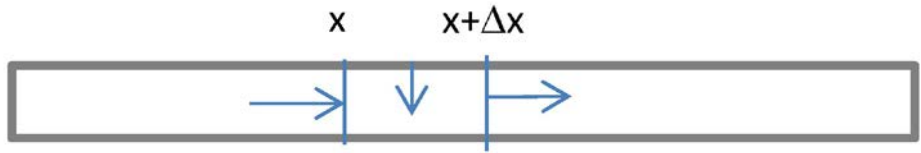

**Fig. 9:** Schematic geometrical model used to write the mass balance equation

$$Q(x+\Delta x) - Q(x) = 2\pi R \Delta x\, q \;\Rightarrow\; \frac{\mathrm{d}Q}{\mathrm{d}x} = 2\pi R q,$$

$$Q(x+\Delta x) = -C\frac{L}{\Delta x}\big(P(x+\Delta x)-P(x)\big) = -CL\frac{\Delta P}{\Delta x} \;\Rightarrow\; Q(x) = -CL\frac{\mathrm{d}P}{\mathrm{d}x},$$

$$\Rightarrow\; CL\frac{\mathrm{d}^2 P}{\mathrm{d}^2 x} = -2\pi R q, \qquad (38)$$

where $q$ is the outgassing rate of the unit surface area, and $C$, $CL$ and $CL/\Delta x$ are the conductances of the beam pipe, the unit length and the section $\Delta x$ of the beam pipe, respectively.

Equation (38) indicates that the pressure profile is parabolic. The analytical expression is obtained by imposing the pressure at the pump location $P(0)$ and the absence of pressure gradient at the opposite beam pipe extremity (see Fig. 10).

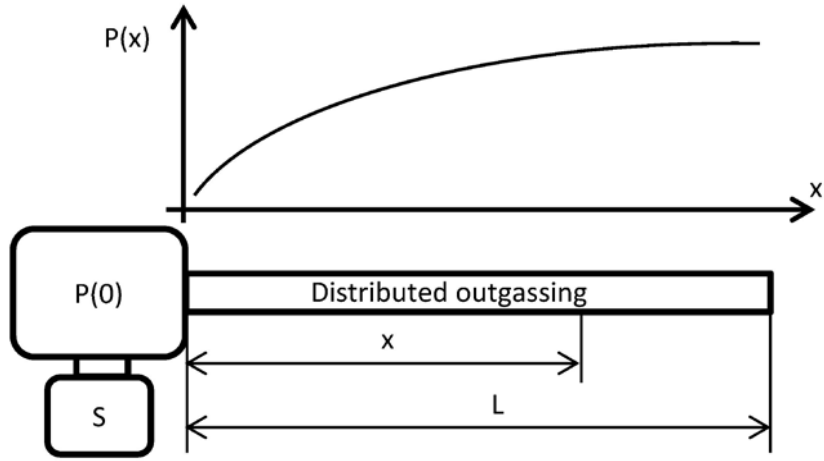

**Fig. 10:** Pressure profile in a tube pumped at one extremity with uniformly distributed outgassing rate

$$P(0) = \frac{Q_{TOT}}{S} = \frac{2\pi RLq}{S},$$

$$\left(\frac{dP}{dx}\right)_{x=L} = 0.$$

Therefore,

$$P(x) - P(0) = -\frac{Q_{TOT}}{C}\left[\left(\frac{x}{L}\right) - \frac{1}{2}\left(\frac{x}{L}\right)^2\right],$$

$$P(L) - P(0) = -\frac{Q_{TOT}}{2C}. \tag{39}$$

where $Q_{TOT}$ is the outgassing rate of the whole beam pipe.

If a second identical pump is installed on the other extremity of the beam pipe (see Fig. 11), the pressure profile is again parabolic, with a maximum at the centre of the vacuum chamber ($x = L/2$):

$$P(0) = \frac{2\pi RLq}{2S} = \frac{Q_{TOT}}{2S},$$

$$P(x) - P(0) = -\frac{Q_{TOT}}{2C}\left[\left(\frac{x}{L}\right) - \left(\frac{x}{L}\right)^2\right], \tag{40}$$

$$P\left(\frac{L}{2}\right) - P(0) = \frac{Q_{TOT}}{8C}.$$

Numerical codes and Monte Carlo simulations are available for geometries that are more complex. Dedicated codes have been extensively used [14].

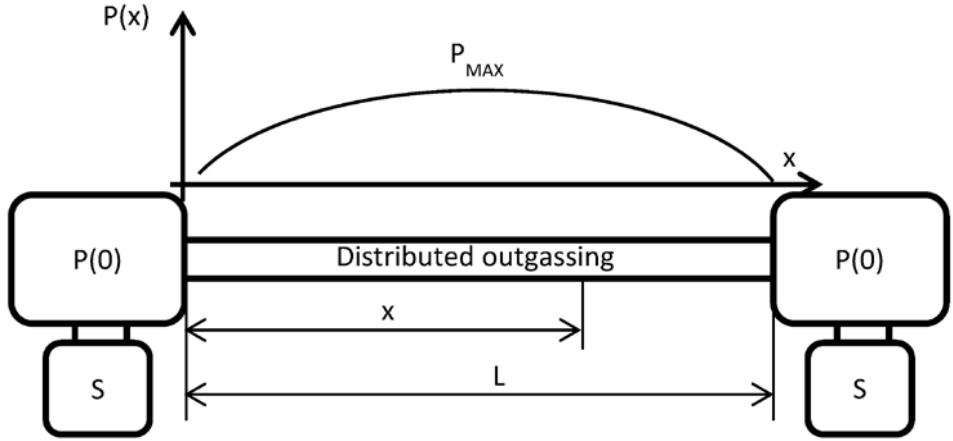

**Fig. 11:** Pressure profile in a tube pumped at both extremities with uniformly distributed outgassing rate

### 3.3 Time-dependent pressure profiles

The calculations shown in the previous subsection are valid only for time-independent conditions. In time-dependent conditions, the variation of the quantity of gas in the vacuum system has to be added to the balance equations such as Eq. (38). For the simplest vacuum system (vessel and pump), the gas balance equation becomes

$$\frac{dN}{dt} = Q - SP,$$

where $Q$ is the gas load, $SP$ is the gas removed by the pump, and $N$ is the number of molecules in the vacuum system. Differentiating Eq. (2) and converting to pressure–volume units, the following equation is obtained:

$$V\frac{dP}{dt} = Q - SP. \tag{41}$$

Equation (41) is easily solved:

$$P(t) = Ae^{-t/\tau_p} + \frac{Q}{S}.$$

The characteristic time of pumping $\tau_p = V/S$ characterizes all pressure transients in a vacuum system. The integration constant $A$ is calculated by imposing the pressure at $t = 0$. An exponential pressure variation is obtained, attaining the stationary value asymptotically. The pressure variations for $P(0) = P_0$ and $P(0) = 0$ are

$$P(0) = P_0 \implies P(t) = \left(P_0 - \frac{Q}{S}\right)e^{-t/\tau_p} + \frac{Q}{S}, \tag{42}$$

$$P(0) = 0 \implies P(t) = \frac{Q}{S}\left(1 - e^{-t/\tau_p}\right). \tag{43}$$

In the case of time-dependent gas sources, as in pulsed ion sources, the solution of Eq. (41) is given by

$$P(t) = \frac{\int e^{t/\tau_p} \frac{Q_{\text{in}}(t)}{V} dt + A}{e^{t/\tau_p}} \tag{44}$$

$$= \frac{\int Q_{\text{in}}(t)dt}{V} - \frac{e^{-t/\tau_p}}{V\tau_p}\int e^{t/\tau_p}\left[\int Q_{\text{in}}(t)dt\right]dt + Ae^{-t/\tau_p}.$$

For complex vacuum systems where several components are interconnected, the balance equation (41) is written for each vessel. This results in a system of coupled differential equations whose solution could be time-consuming. However, the analogy between vacuum systems and electrical networks may be used to accelerate the calculation.

### 3.4 Calculation of pressure profiles by the electrical analogy

Equations (11), (26) and (27) show that there is an analogy between vacuum systems and electrical networks. In Table 11 vacuum components and variables are correlated with electrical elements and characteristics.

**Table 11:** Electrical analogy of vacuum components and variable.

| Vacuum element | Electrical elements | Electrical symbol |
|---|---|---|
| Conductance $C$ | Conductance $1/R$ | 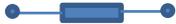 |
| Gas flow $Q$ | Current $I$ | |
| Pressure $P$ | Voltage $V$ | |
| Volume $V$ | Capacitance $C$ | 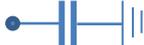 |
| Pump | Conductance to ground | 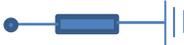 |
| Gas source | Current generator | 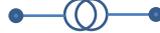 |
| Constant pressure source | Voltage supply | 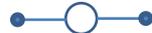 |

The ground potential is equivalent to zero pressure. A vacuum chamber of a given conductance and volume corresponds to two resistances and a capacitance. For symmetry, the capacitance is placed in the middle of the two resistances. If a local gas source and a pump are added, a current generator and a resistance to ground are connected to the circuit (see Fig. 12).

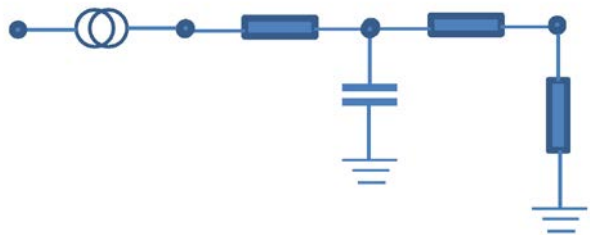

**Fig. 12:** Electrical analogy of a vacuum chamber with gas injection and pumping

Another typical example is given in Fig. 13; it represents a differential pumping. This configuration is used when a high gas flow must be prevented from reaching the low-pressure side of a vacuum system. It consists of a small diaphragm with pumps on both sides.

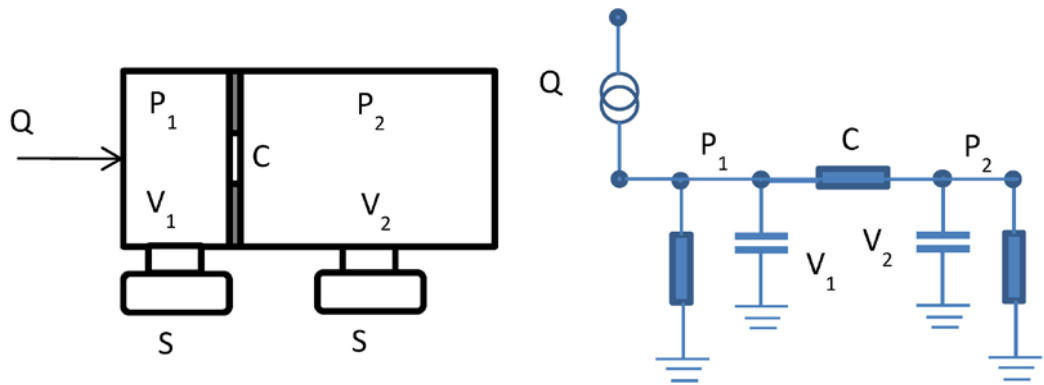

**Fig. 13:** Electrical analogy of a vacuum system with differential pumping

Long beam pipes are subdivided into small units to calculate the axial pressure distribution; the small units are considered as single vacuum chambers (volume and conductance) in series. The conductance of a single small unit is equal to the conductance of the entire vacuum chamber times the number of units. Distributed outgassing is taken into account by inserting a current generator on each unit.

The electrical network is solved by dedicated software, for example LTSpice. The time evolution and pulsed sources are easily included in the calculation. Nonlinear electrical components are used to simulate pressure- and time-dependent conductances and pumping speeds.

The MedAustron ring [15] and the Linac4 ion source [16] have been fully simulated by means of the electrical analogy. The effects of modifications in the shape of the components, position and size of the pumps have been checked and optimized.

## 4    Gas pumping

Vacuum pumps in the molecular regime are classified into two families: momentum transfer pumps and capture pumps. Both act on each molecule singularly since no momentum and energy transfer is possible between molecules in this pressure range. In the first family, molecules receive a momentum component pointing towards the pump outlet (foreline) where the gas is compressed and evacuated by pumps working in the viscous regime (e.g. rotary vane, diaphragm and scroll pumps). The second family removes gas molecules by fixing them on a surface exposed to the vacuum.

### 4.1    Momentum transfer pumps

In this family of pumps, gas molecules are steered by collision with moving surfaces (molecular pumps) or supersonic jets (diffusion pumps). The latter mechanism is no longer used in particle accelerators, though it still has extensive applications in industrial vacuum systems.

#### *4.1.1    Molecular pumps*

In molecular pumps, gas molecules impinge and adsorb on the moving surface; on desorption, the molecular emission is no longer isotropic due to the superposition of the wall velocity (see Fig. 14). As a result, molecules are preferentially redirected towards the direction of the wall movement, i.e. the density of molecules increases in the same direction.

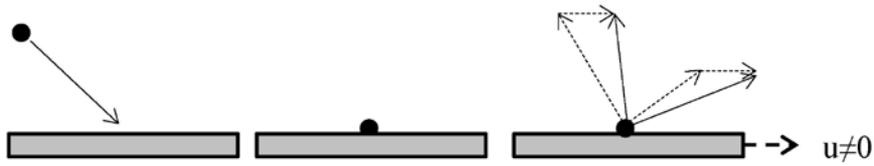

**Fig. 14**: Schematic drawing showing the mechanism of momentum transfer by collision onto a moving surface

The first molecular pump was invented by W. Gaede. In the original design, the moving surface is a rotor revolving at high frequency (see Fig. 15). To prevent back-streaming, the inlet and the outlet were separated by a very thin slot (locking slot). The slot height was of the order of $10^{-2}$ mm.

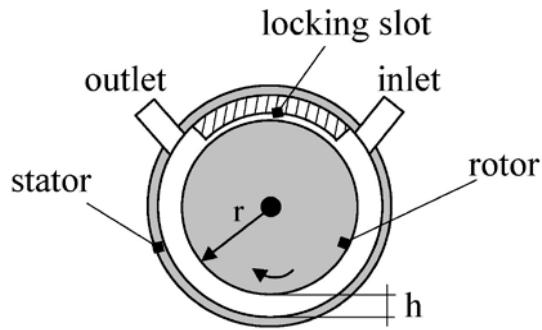

**Fig. 15:** Schematic drawing of a molecular pump

The most important characteristics of a molecular pump are the pumping speed $S$ and the maximum compression pressure ratio that can be achieved between the pump inlet $P_{IN}$ and outlet $P_{OUT}$:

$$K_0 = \left(\frac{P_{OUT}}{P_{IN}}\right)_{MAX}. \tag{45}$$

The crucial parameters affecting $S$ and $K_0$ are highlighted by a simplified model (see Fig. 16).

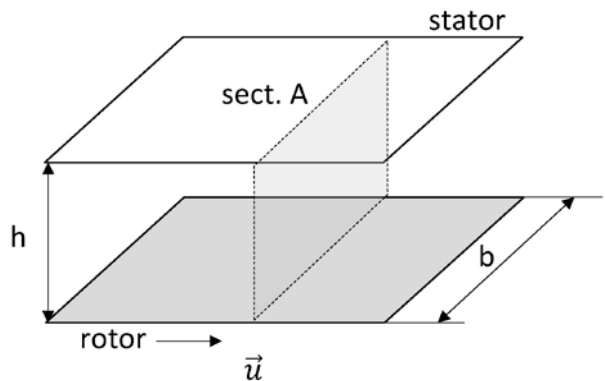

**Fig. 16:** Plane section of a molecular pump

At any time, half of the molecules have just collided with the moving surface (rotor) and drift in the direction of the velocity $\vec{u}$. The other half hit the stator where the drift component is lost. As a result, the drifted molecular flow $Q_p$ towards an imaginary cross-sectional area $A$ (see Fig. 16) is

$$Q_\text{p} = \frac{1}{2} nubh = \frac{1}{2} \frac{P}{k_\text{B} T} ubh,$$

where '*bh*' is the area *A* of the cross-section and '*ubh*' is the gas volume drifting through area *A* in one second. Converting into pressure–volume units and dividing by *P*, the pumping speed is obtained as

$$S = \frac{Q_p}{P} = \frac{1}{2} ubh. \tag{46}$$

Therefore, the pumping speed of molecular pumps is proportional to the speed of the moving wall. It does not depend on the nature of the gas; thus, molecular pumps are not selective. Finally, a large cross-section '*bh*' is needed to receive a large quantity of gas.

The moving wall produces a higher gas density near the pump outlet. In turn, the pressure gradient generates a gas backflow $Q_\text{bf}$ that can be written as

$$Q_\text{bf} = -\bar{c}\, \nabla P,$$

where $\bar{c}$ is the conductance of a unit-length duct of cross-sectional area *bh*. For $h \ll b$, one has

$$\bar{c} = \frac{2}{3} \frac{(bh)^2}{b+h} \langle v \rangle.$$

The net flow is therefore given by

$$Q = Q_\text{p} - Q_\text{bf} = \frac{1}{2} Pubh - \bar{c}\, \nabla P.$$

When the net flux is zero, the maximum compression ratio is reached, in which case

$$\frac{dP}{P} = \frac{1}{2} \frac{ubh}{\bar{c}} dx.$$

Integrating between inlet and outlet (distance *L*) and using Eqs. (13) and (20) gives

$$K_0 = \left(\frac{P_\text{OUT}}{P_\text{IN}}\right)_\text{MAX} = \exp\left(\frac{ubhL}{\bar{c}}\right) \propto \exp\left(\frac{u}{\langle v \rangle} \frac{L}{h}\right) \propto \exp\left(u\sqrt{m_i}\, \frac{L}{h}\right). \tag{47}$$

To obtain high $K_0$:

– the velocity of the moving surface must be of the order of the mean molecular speed (high $u/\langle v \rangle$ ratio), and
– the duct where molecules drift must be narrow and long (high $L/h$),

In addition, $K_0$ depends strongly on the molecular mass, the lowest being for $H_2$. As a result, the ultimate pressure of molecular pumps is dominated by $H_2$. More generally, the low $K_0$ for $H_2$ is an intrinsic limitation of momentum transfer pumps.

### 4.1.2  Turbomolecular pumps

For many years, the industrialization of molecular pumps was hindered by the machining of the locking slot. The narrow gap between rotor and stator created mechanical issues due to tolerances and thermal expansion. The problem was overcome in 1957 when W. Becker invented the turbomolecular pump (TMP). In the TMP, the momentum transfer is produced by rapidly rotating blades rather than parallel surfaces.

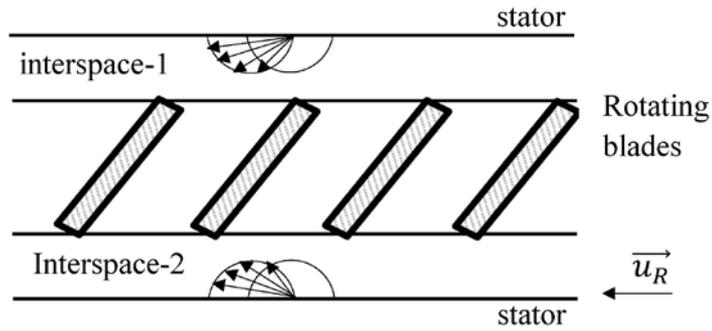

**Fig. 17:** A single rotor–stator stage in a turbomolecular pump, showing the molecular speed distributions

The rotating blades are tilted with respect to the rotational axis. As a result, oblique channels between successive blades are formed. The rotor is inserted between two static surfaces defining two interspaces (see Fig. 17). When the molecule comes from interspace-1, the angular distribution of the molecular velocity seen from the blades is preferentially oriented towards the blades' channels. Conversely, for those coming from interspace-2, the velocity vectors point preferentially towards the blades' wall, therefore increasing backscattering. A gas flow is consequently generated from interspace-1 to interspace-2. This mechanism works only if the angular distribution of the molecular speed as seen from the blade is significantly deformed, i.e., only if the blades' speed is at least of the order of the mean molecular speeds (hundreds of metres per second).

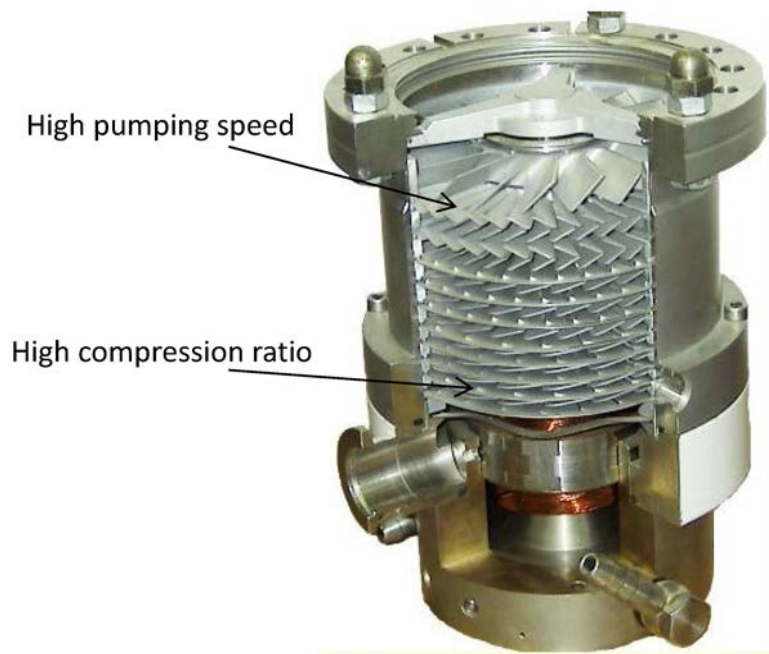

**Fig. 18:** Cut-out view of a turbomolecular pump. Courtesy of Wikipedia (http://en.wikipedia.org/wiki/Turbomolecular_pump).

In a real turbomolecular pump, the gas is compressed by several series of rotating blades (see Fig. 18). Every series of rotating blades is followed by a series of static blades. Molecules transmitted through the rotating blades' channels hit the static blades; as a result, the angular distribution of velocity is randomized and the molecules are ready for the next compression stage. The momentum transfer is effective only if the molecules do not experience intermolecular collisions after hitting the blades; this is equivalent to saying that the mean free path has to be larger than the blade distances. As a result, this type of pump works at full pumping speed only in molecular regimes ($P < 10^{-3}$ mbar).

The conclusions drawn for molecular pumps hold also for turbomolecular pumps. The dimension of the blades and the rotor–stator distance are larger for the first series of blades in view of maximizing the pumping speed (see Eq. (46)). The last compression stages are tighter to increase the maximum compression ratio (see Eq. (47)) and compensate for the increased gas density and thus for the resulting lower molecular mean free path. The pumping speed of a TMP (Fig. 19) is constant in the molecular regime ($P < 10^{-3}$ mbar) and ranges between 10 and 3000 l s$^{-1}$ depending on pump inlet diameter and mechanical design. As expected, the maximum compression ratio is the lowest for $H_2$; in classical designs it is about $10^3$ (see Fig. 20). Nowadays, values up to $10^6$ are achieved by integrating a molecular drag pump (Gaede or Holweck drag stage) to the rear of the blade sets.

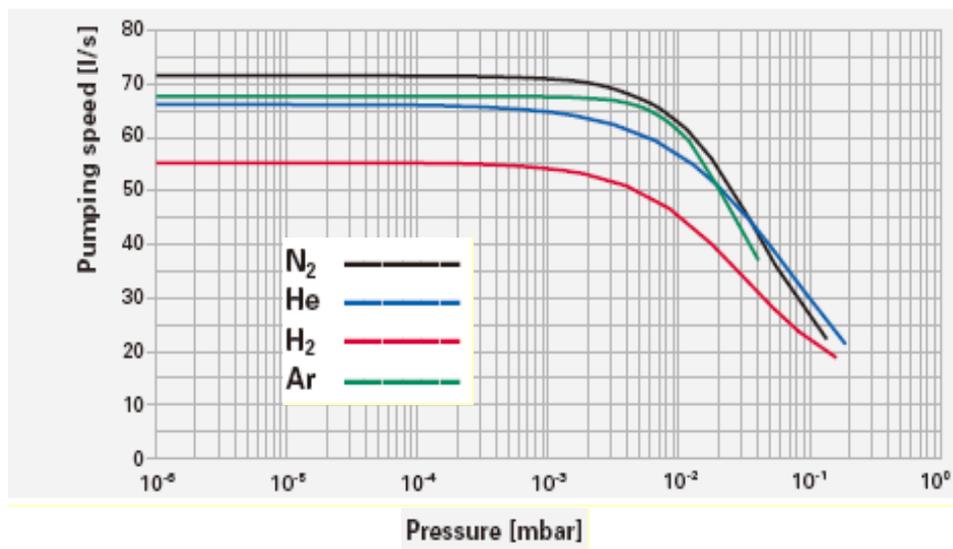

**Fig. 19**: Pumping speeds of a turbomolecular pump equipped with a DN63CF flange. For TMP, the pumping speeds are constant in the free molecular regime and the nature of the pumped gas has a limited effect, if compared with other type of pumps. Courtesy of Pfeiffer Vacuum.

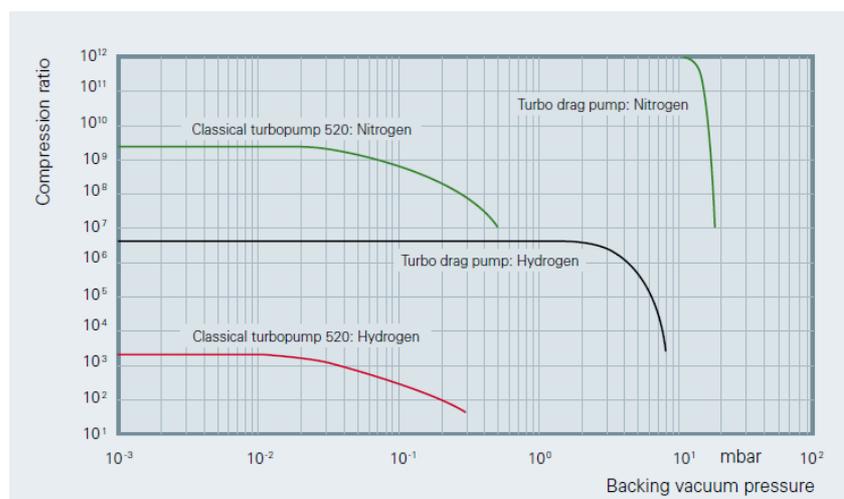

**Fig. 20:** Maximum compression ratio of two different turbomolecular pumps for $N_2$ and $H_2$. Courtesy of Pfeiffer Vacuum.

The TMP ultimate pressure is of the order of $10^{-10}$–$10^{-11}$ mbar for baked and all-metal vacuum systems. Lower pressures may be obtained if another type of pump removes the TMP $H_2$ back-streaming.

At the pump outlet, the compressed gas is evacuated by a mechanical pump (often called backing pump) operating in the viscous regime. The turbomolecular pump and its backing pump are in general assembled in a single unit that includes power supplies, controls and instrumentation.

Each set of rotor blades is machined from a single block of high-strength aluminium alloys. Typical pumps equipped with a DN100 flange rotate at a frequency of 1 kHz, therefore approaching circumferential speeds up to 500 m s$^{-1}$.

The TMP is the best available pump for the first stages of ion sources. It can evacuate high gas flow at relatively high pressure without selectivity and memory effects. For example, a DN100 TMP can withstand a continuous gas flow up to $2 \times 10^{-1}$ mbar l s$^{-1}$ without damage.

The use of dry pumps for TMP backing has surmounted the risk of oil back-streaming from rotary vane pumps. The higher ultimate pressure of dry pumps is compensated by the integration of molecular drag pumps (higher compression ratio). The complete removal of all lubricated mechanical bearings has been obtained by magnetic rotor suspension.

The main drawback of a TMP is related to possible mechanical failures leading to definitive damage of the high-speed rotor. In addition, in case of unwanted rotor deceleration caused by power cut or rotor seizing, the vacuum system has to be protected by safety valves and dedicated pressure sensors against air back-stream. TMPs have limited application in radioactive environments due to possible damage to their electronics and power supplies. Some radiation-resistant TMPs are now available; with a significant cost, the whole electronics is moved beyond the radiation shielding by means of long cables.

### 4.2 Capture pumps

Capture pumps remove gas molecules by fixing them onto an internal wall. To be efficient, capture pumps must block the gas molecules for an average time (sojourn time $t_s$) much longer than the typical running time of the accelerator. An estimation of the sojourn time is given by the Frenkel law [17]:

$$t_s = t_0 \, e^{E_a/k_B T}, \tag{48}$$

where $E_a$ is the adsorption energy and $t_0 \approx h/k_B T \approx 10^{-13}$ s. Very long sojourn times are obtained either for high adsorption energies

$$E_a \gg k_B T,$$

or for very low temperatures

$$T \ll \frac{E_a}{k_B}.$$

In the former case, capture pumps are called *chemical pumps* or *getter pumps*; in the latter case, *cryogenic pumps*.

In chemical pumps, the binding force involves electron exchange between gas molecules and surfaces (*chemisorption*). Typical binding energies for getter pumps are higher than 1 eV/molecule. In cryogenic pumps, Van der Waals' interactions might be sufficiently strong to fix molecules to cold surfaces (*physisorption*). In this case, the binding energy can be much lower than 0.5 eV/molecule.

Chemisorption, implantation and physical burial of gas molecules by reactive metal atoms are combined in sputter ion pumps. They represent the most important and widely used pumping technology in particle accelerators.

### 4.2.1 *Sputter ion pumps*

In a sputter ion pump (SIP) the residual gas is ionized in a Penning cell. The ions are then accelerated towards a cathode made of a reactive metal (Fig. 21). The cathode–ion collisions provoke sputtering of reactive-metal atoms that are deposited on the nearby surfaces (Fig. 22). The Penning cells are assembled in honeycomb structures to form the pump's elements.

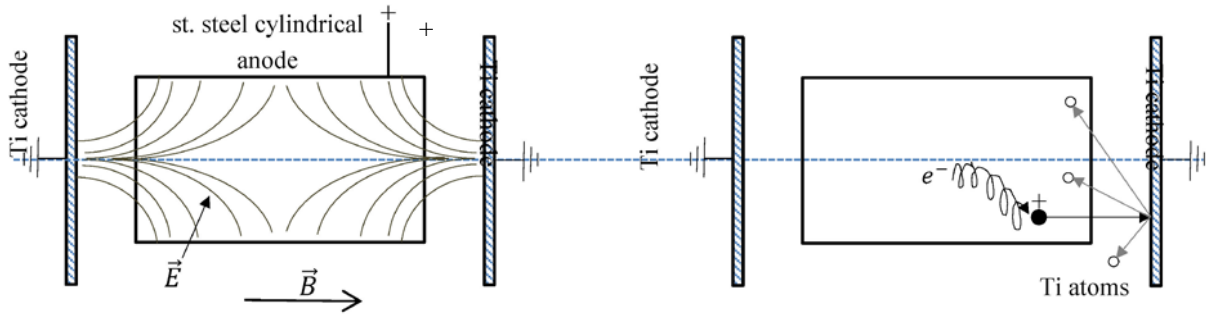

**Fig. 21:** Schematic drawing depicting the pumping mechanism of sputter ion pumps in the diode configuration

The pumping action is given by three distinct mechanisms:

– Chemical adsorption of gas molecules onto the reactive metal layer and subsequent burial by additional sputtered atoms; this works for all gas species except for rare gases, which cannot form chemical compounds.

– Implantation both of gas ions in the cathode and of energetic neutrals bounced back from the cathode into the deposited film; this is the only pumping mechanism for rare gases.

– Diffusion into the cathode material and the deposited film; this is possible only for $H_2$ because of its very large diffusivity in metals in its atomic form.

In the diode SIP (Fig. 21), the anode is composed of several open cylinders made of stainless steel at a positive electrical potential (from 3 to 7 kV). The cathodes are plates of Ti – at ground potential – placed a few millimetres from both anode extremities. A magnetic field parallel to the cells' axis is generated by external permanent magnets (about 0.1 T). In this configuration, the crossed electrical and magnetic fields trap electrons in long helical trajectories, resulting in an increased probability of gas ionization.

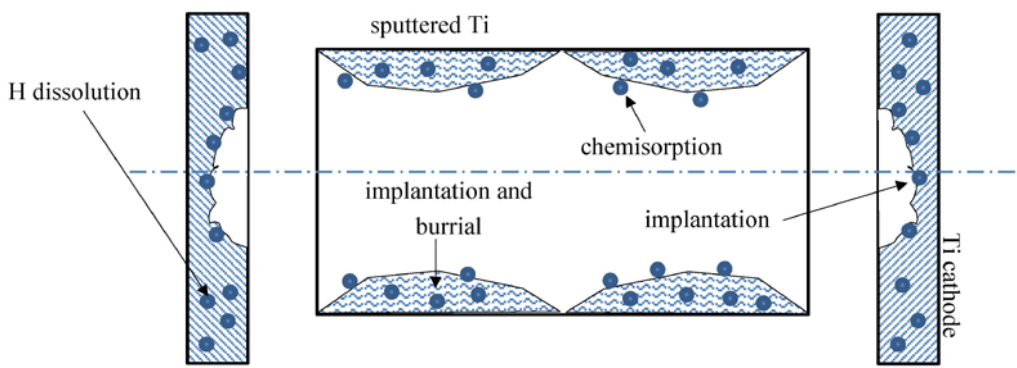

**Fig. 22:** Schematic drawing of a SIP cell after prolonged operation

Molecular implantation in the cathodes is not a permanent pumping; in fact, the progressive erosion due to gas ion sputtering sooner or later frees the implanted gas (Fig. 22). Once released from the cathodes, reactive gas species are chemisorbed while rare gases remain in the gas phase until the next implantation. Pressure instabilities have been reported after excessive pumping of rare gas: the continual erosion liberates gas that increases the pressure, which in turn increases the erosion. Consequently, a pressure rise is generated; this is stopped when most of the rare gas is implanted in a deeper zone of the cathode. A new pressure spike appears when the erosion reaches the new rare gas front. This type of instability is typically observed when large quantities of air are pumped (Ar represents 1% in air) and is known as 'argon disease' [18].

The pumping efficiency for rare gas is improved by reducing the number of ions implanted into the cathodes, while increasing the energetic neutrals bounced back from the cathode and their chance to be buried by Ti atoms onto the anode. Two different approaches have been commercially exploited:

–  cathodes made of metals with higher atomic mass, and
–  different geometry than the diode configuration.

In the first case, Ta is used instead of Ti (181 amu against 48 amu). Gas ions, once neutralized on the cathode surface, have a higher probability to escape from the cathode and bounce back with higher energy when colliding on Ta rather than Ti. Consequently, the implantation probability on the anode is higher and the quantity of gas implanted into the cathode is lower. Pumps that use Ta cathodes are called 'noble diode' or 'differential ion' pumps.

In the second case, the diode configuration is modified by introducing a third electrode (triode sputter ion pumps, see Fig. 23). The cathode consists of a series of small Ti plates aligned along the cell axis and polarized negatively with respect to the ground. The anode is at the ground potential.

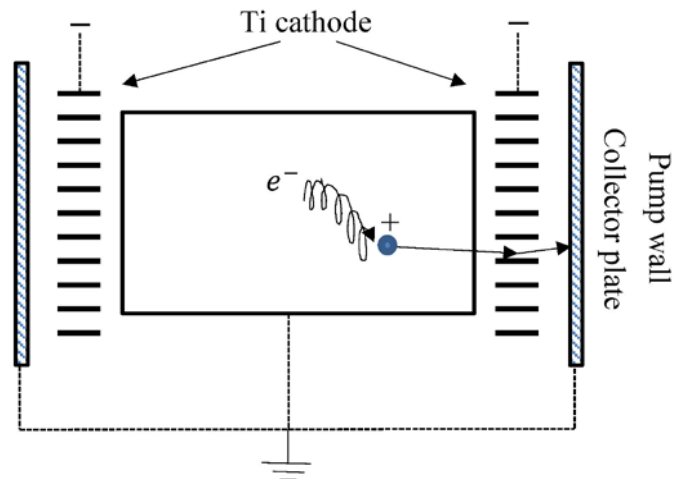

**Fig. 23:** Schematic drawing of a triode SIP cell. The Ti cathodes are negatively biased

In this configuration, the collisions between ions and cathode are at glancing angle: the sputtering rate, the neutralization and the bouncing probability are higher. As a result, more Ti atoms are sputtered and more gas molecules are implanted into the collector plate and anode; the gas implanted into the cathode is minimized.

An improved triode pump is the StarCell® produced by Agilent. Two sheets of Ti that are cut and bent to form radial blades (see Fig. 24) replace the series of small plates. This increases the rigidity of the cathode and reduces the risk of short-circuits.

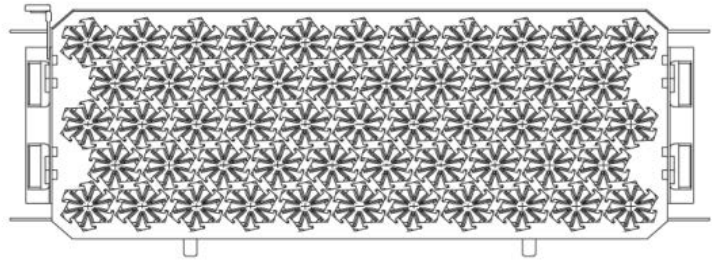

**Fig. 24:** Drawing of Agilent's StarCell Ti cathode. Courtesy of Agilent Vacuum

*4.2.1.1  Hydrogen pumping by SIP*

In SIP, $H_2$ is mainly pumped by dissolution into the Ti cathodes. To be absorbed, $H_2$ must be dissociated. Because in the Penning discharge only a small fraction of the hydrogen ions are $H^+$, the dissociation is possible only on the Ti cathodes after removal of the oxide layer by sputtering. This surface cleaning process is very slow if $H_2$ is the leading gas in the vacuum system due to its low sputtering yield (0.01 at 7 keV on Ti). Therefore, at the beginning of the operation, the pumping speed for $H_2$ is lower than the nominal value and increases gradually with time.

The simultaneous pumping of another gas has a strong effect on $H_2$ pumping efficiency. If the additional gas has a higher sputtering yield, the cathode cleaning is faster and the nominal pumping speed is reached in a shorter time. Conversely, it can contaminate the cathode surface and reduce the $H_2$ adsorption. Finally, dissolved H atoms can be sputtered away, reducing the pumping efficiency.

When the concentration of $H_2$ in the cathodes is higher than the solubility limit in Ti, hydride precipitates are formed. The cathodes expand and become brittle, generating anomalous field emission and short-circuits. A typical concentration limit is 10 000 mbar l for a commercial 500 l s$^{-1}$ SIP. Water vapour pumping also contributes to the total quantity of $H_2$ charged in the cathodes.

During the pumping of $H_2$ or after the dissolution of a high quantity of the same gas, the operation of SIP at high pressures (higher than $10^{-5}$ mbar) can lead to thermal run-away. The Penning discharge heats the cathode and provokes gas desorption, which reinforces the discharge. This positive feedback mechanism can cause local melting of the cathode. To avoid this problem, the electrical power provided to the pump is limited at high pressure and during the ignition.

*4.2.1.2  Pressure measurement by SIP*

The ion current collected by the SIP's cathodes is used as an indication of the pressure. In fact, a defined relationship is measured between the two variables down to pressures of the order of $10^{-9}$ mbar. For lower pressures, the pump current is limited by field emission (leakage current). The low-pressure reading can be extended by one decade if the applied voltage is reduced in the lower-pressure range (e.g. from 7 to 3 kV). The pressure reading by SIP is extensively used in particle accelerators.

*4.2.1.2  Pumping speed of SIP*

The pumping speed of SIP depends on the gas pressure at the pump inlet, the nature of the gas, and the quantity of gas already pumped. The latter dependence is correlated with the implantation into the cathode. For new pumps and gases other than $H_2$, the rate of implantation is larger than the rate of release by sputtering. Conversely, for 'saturated' pumps, the two rates equate and the pumping in the cathode is negligible. The nominal pumping speed of SIP is in general given for 'saturated' pumps.

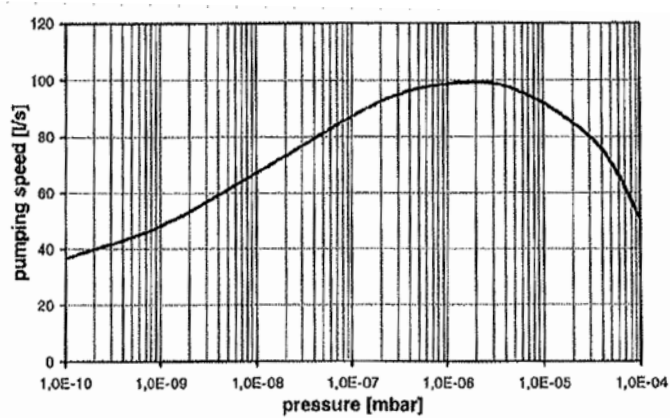

**Fig. 25:** Typical inlet pressure dependence of SIP pumping speed; the maximum pumping speed is reached at about $10^{-6}$ mbar. Courtesy of Agilent Vacuum.

The pumping speed attains a maximum for inlet pressures of about $10^{-6}$ mbar. This is the value given by the suppliers for the nominal pumping speed. For lower pressures, the pumping speed decreases, half of the maximum being obtained at about $10^{-9}$ mbar (see Fig. 25). Typical values of $S$ are collected in Tables 12 and 13. A cut-away view of the StarCell pump is shown in Fig. 26.

**Table 12:** $N_2$ nominal pumping speed of commercial StarCell SIP ('saturated') for different standard pump inlet diameters.

| Pump inlet diameter, DN (mm) | $S_{N_2}$ (l s$^{-1}$) |
|---|---|
| 63 | 50 |
| 100 | 70 and 125 |
| 150 | 240 and 500 |

**Table 13:** Nominal pumping speed normalized to that of air for diode and triode SIP.

| Gas | Air | $N_2$ | $O_2$ | $H_2$ | CO | $CO_2$ | $H_2O$ | $CH_4$ | Ar | He |
|---|---|---|---|---|---|---|---|---|---|---|
| Diode | 1 | 1 | 1 | 1.5-2 | 0.9 | 0.9 | 0.8 | 0.6-1 | 0.03 | 0.1 |
| Triode | 1 | 1 | 1 | 1.5-2 | 0.9 | 0.9 | 0.8 | 0.6-1 | 0.25 | 0.3 |

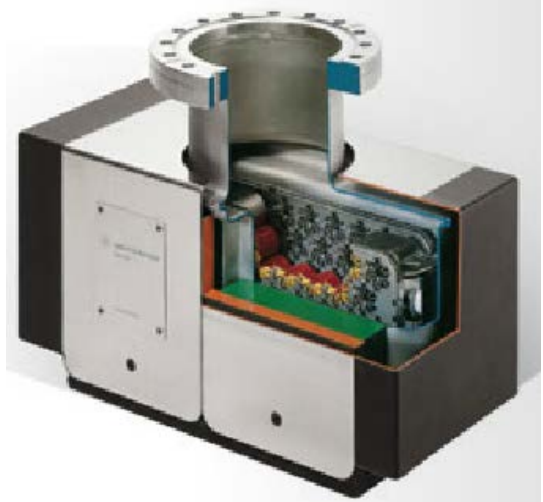

**Fig. 26:** Cut-away view of a triode StarCell pump. The inlet flange is DN63CF. Courtesy of Agilent Vacuum

*4.2.2    Getter pumps*

Getter materials fix gas molecules on their surface by forming stable chemical compounds. The chemical reaction is possible only if the getter atoms are not already combined with surface contaminants, i.e. oxygen, carbon, etc. A contamination-free surface may be produced in two ways:

– sublimating the getter material *in situ* – evaporable getters or sublimation pumps; and

– dissolving the surface contamination into the bulk of the getter material by heating – non-evaporable getters (NEGs); the dissolution process is called activation.

A getter surface is characterized by its sticking probability $\alpha$, namely the capture probability per molecular impingement. From Eq. (32), the pumping speed of getter material is written as

$$S = \alpha A_{\text{getter}} C', \qquad (49)$$

where $A_{\text{getter}}$ is the geometric surface area of the getter material. The $\alpha$ value depends on many variables, first of all the nature of the gas. Getter materials do not pump rare gases and methane at room temperature. In that respect, they always need auxiliary pumping to keep a stable pressure. The sticking probabilities of homonuclear diatomic molecules are in general lower than those of molecules composed of dissimilar atoms. The values of $\alpha$ for common gas species in vacuum systems may be placed in increasing order as follows:

$$N_2 < H_2 < O_2 < H_2O < CO < CO_2.$$

For $N_2$ typical $\alpha$ values are in the high $10^{-3}$ to low $10^{-2}$ range; for CO and $CO_2$, $\alpha$ values close to 1 may be measured. In addition, the sticking probability is strongly dependent on the gas surface coverage, i.e. the quantity of gas previously pumped on the getter material. As the gas surface coverage increases, the sticking probability decreases. The pumping becomes negligible when the surface is fully covered by gas molecules. As an example, complete saturation of a smooth NEG surface is obtained when about $10^{15}$ CO molecules $cm^{-2}$ are adsorbed. Quantities about 5–10 times higher are reported for $H_2O$ and $O_2$. $H_2$ dissociates on the getter surface and diffuses into the bulk of the NEG material. As a result, its pumping does not block the adsorption of other molecules. The pumping capacity of hydrogen is orders of magnitude higher than that for other gas species.

The surface roughness also has a strong impact on the sticking probability. A single molecular arrival may result in several collisions with a rough getter surface, therefore enhancing the capture probability. The pumping capacity per unit geometric area is also increased.

Owing to the high mobility of H atoms and the relatively low binding energy, the pumping of this gas is reversible: $H_2$ may be released by heating. The opposite effects of pumping and desorption set up an equilibrium pressure $P_{H_2}$ that depends on heating temperature $T$ and hydrogen concentration $c_H$. The three variables are correlated by Sieverts' law, i.e. the law of mass action for the $H_2$–metal system:

$$\log P_{H_2} = A + 2\log c_H - \frac{B}{T} \tag{50}$$

where $A$ and $B$ are typical values for a given getter material.

### 4.2.2.1 Sublimation pumps

In particle accelerators, Ti is the only getter material used as evaporable getter. It is sublimated on the inner walls of a dedicated vessel from Ti–Mo filaments heated up to 1500°C (Fig. 27). The maximum sticking probability varies in the range $(1–5) \times 10^{-2}$ for H2 and 0.5–1 for CO (Fig. 28). In some cases, the wall of the sublimation pumps may be cooled to liquid-nitrogen temperature. In this case, the sublimated film is rougher and consequently the sticking probabilities are higher. When the deposited film is saturated, a new sublimation is needed to recover the initial pumping speed. A single filament withstands tens of sublimation cycles.

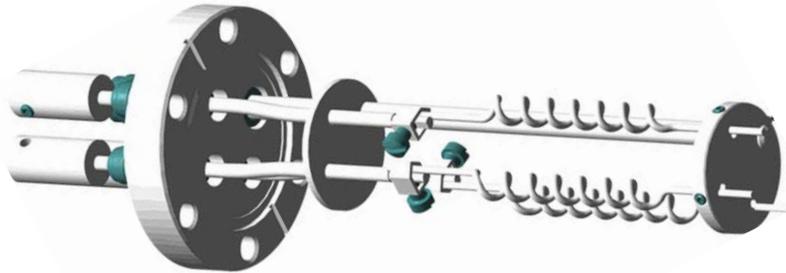

**Fig. 27:** Ti wires of a sublimation pump. Courtesy of Kurt J. Lesker Company

Sublimation pumps provide very high pumping speeds for a relatively low cost and have a limited maintenance. They are in general used to achieve pressures in the UHV range. Operation with continuous sublimation in higher pressure ranges may be conceived; in this case, cooling of the pump envelope must be added.

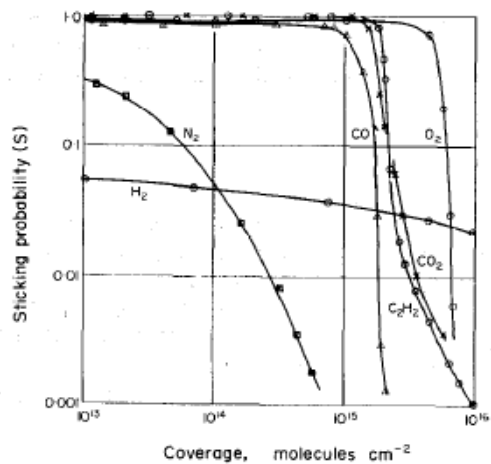

**Fig. 28:** Typical single-gas coverage dependence of the pumping speed of sublimation pumps for selected gas species. This plot and other interesting measurements with sublimation pumps may be found in the very instructive paper of Ref [19].

*4.2.2.2   Non-evaporable getter pumps*

During the activation process, the native oxide layer is dissolved in the NEG material. This relocation is possible only if it is energetically feasible and the bulk of the material has a large solubility limit for oxygen. Only few elements have these peculiar characteristics. Among them, those of the fourth group (Ti, Zr and Hf) are the most relevant. These metals are able to dissolve oxygen at room temperature to more than 20% in atomic concentration. Other metallic elements are added to increase oxygen diffusivity and so reduce the activation temperature and time; this is typically the case for V. For specific purposes, other elements may be added in the alloy, for example to reduce pyrophoricity in air.

NEG materials are usually produced in powder form to increase the surface exposed and the pumping capacity. The powdered materials are either sintered to form discs and pellets or pressed on metallic substrates shaped as filaments and ribbons. The latter have found an extensive application in particle accelerators as linearly distributed pumps [20] (see Fig. 29). Thin-film NEG coatings produced by magnetron sputtering are also used in modern accelerators in the UHV pressure range [21].

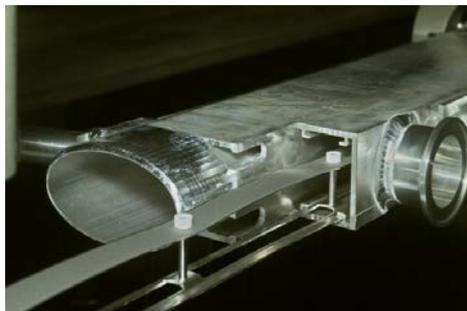

**Fig. 29:** Photograph of the dipole magnet vacuum chamber of the Large Electron–Positron collider. The NEG pump is made of Zr–Al powder fixed on both sides of a constantan ribbon.

A typical NEG alloy commercialized by SAES Getters (Milan, Italy) is the St707 made of Zr–V–Fe. It is activated at 400°C for 1 h or at 300°C for about 24 h; such characteristics allow passive activation during the bake-out of the stainless-steel vacuum chamber.

Modern lump NEG pumps provide gas surface capacities about 200 times higher than smooth surfaces. The embrittlement of the NEG elements limits the maximum quantity of $H_2$ that may be dissolved in the pumps. A safe limit of 20 Torr l $g^{-1}$ is given by SAES Getters for the St707. For the same alloy, the $H_2$ equilibrium pressure is shown in Fig. 30 for various temperatures. The constants $A$ and $B$ of Eq. (50) are 4.8 and 6116, respectively, when the pressure is expressed in Torr and the concentration in Torr l $g^{-1}$.

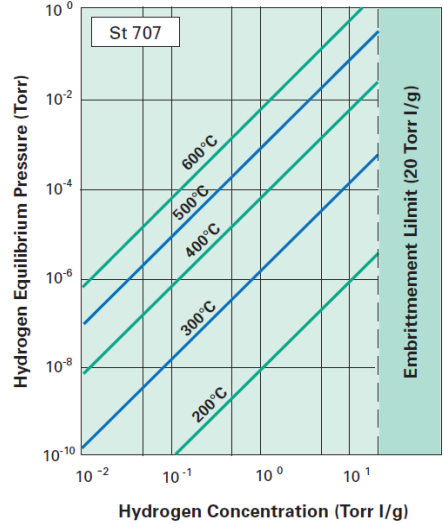

**Fig. 30:** Hydrogen equilibrium pressure for the St707 alloy. Courtesy of SAES Getters

After the pumping of a large quantity of $H_2$, a regeneration of the NEG pump may be necessary to avoid embrittlement. To do so, the NEG material is heated and the released gas is removed by a turbomolecular pump. The duration of the regeneration ($t_R$) is given [22] by

$$t_R = \frac{M_{NEG}}{S_{TMP}} \left( \frac{1}{q_f} - \frac{1}{q_i} \right) \times 10^{-A-(B/T)}, \tag{51}$$

where $M_{NEG}$ (g) is the mass of the NEG material, $S_{TMP}$ (l $s^{-1}$) is the pumping speed of the turbomolecular pump, $q_i$ and $q_f$ (Torr l $g^{-1}$) are the initial and the final concentrations of hydrogen, and $T$ (K) is the absolute temperature.

At room temperature, the pumping speed decreases as a function of the amount of gas pumped, except for $H_2$, which diffuses into the alloy. The drop of pumping speed is less pronounced than for the sublimation pumps, due to the high porosity and related surface area of commercial NEG materials (see Fig. 31). The SAES Getters CapaciTorr pumps (see Fig. 32) have a sorption capacity for CO at room temperature up to about 5 Torr l (~$1.5 \times 10^{20}$ molecules). NEG pumps may operate at higher temperatures; the diffusion of C and O atoms is enhanced and the surface capacities largely increased. The practical result is a less pronounced pumping speed drop as a function of the quantity of gas pumped. However, the $H_2$ release might be an obstacle to achieve the required pressure.

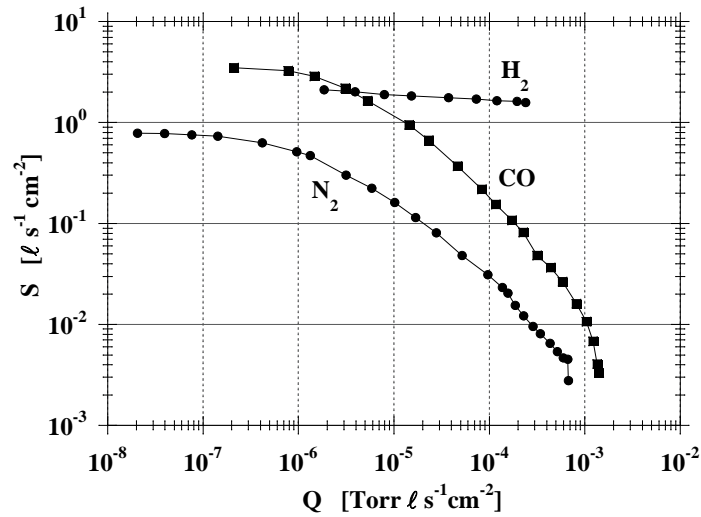

**Fig. 31:** Typical single-gas coverage dependence of the pumping speed of a St707 ribbon for selected gas species. This plot and other interesting measurements with NEG ribbons may be found in [23].

CO and $CO_2$ pumping hinder the adsorption of all other gases, while $N_2$ has only a limited effect and $H_2$ leaves the NEG surface unaffected.

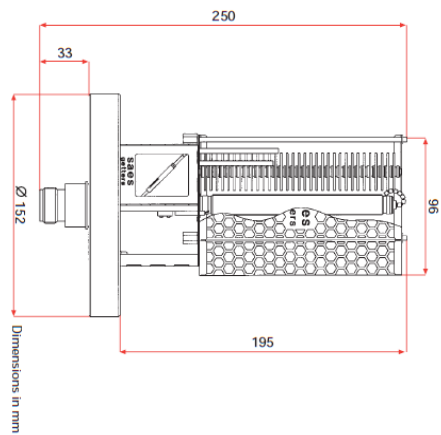

**Fig. 32:** Drawing and dimensions of a CapaciTorr pump. Courtesy of SAES Getters

### 4.2.3 Cryopumps

Cryopumps rely on three different pumping mechanisms:

– *Cryocondensation.* This mechanism is based on the mutual attraction of similar molecules at low temperature (Fig. 33). The key property is the saturated vapour pressure $P_v$, i.e. the pressure of the gas phase in equilibrium with the condensate at a given temperature [24]. The lowest pressure attainable by cryocondensation pumps is limited by the saturated vapour pressure. Among all gas species, only Ne, $H_2$ and He have $P_v$ higher than $10^{-11}$ Torr at 20 K. The $P_v$ of $H_2$ at the liquid He boiling temperature is in the $10^{-7}$ Torr range, and is $10^{-12}$ Torr at 1.9 K. The quantity of gas that may be cryocondensed is very large and limited only by the thermal conductivity of the condensate.

– *Cryosorption.* This is based on the attraction between gas molecules and substrates (Fig. 33). The interaction forces with the substrate are much stronger than those between similar molecules. As a result, providing the adsorbed quantity is lower than one monolayer, the sojourn time is much longer and gas molecules are pumped at pressure much lower than the saturated vapour pressure. A significant quantity of gas may be pumped below one monolayer if porous materials are used; for example, in one gram of standard charcoal for cryogenic application, about 1000 m$^2$ of surface are available for adsorption. The important consequence is that significant quantities of $H_2$ and He may be pumped at 20 K and 4.3 K, respectively. In general, submonolayer quantities of all gas species are effectively cryosorbed at their own boiling temperature.

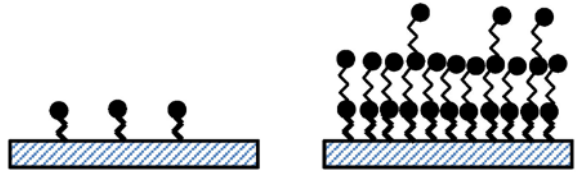

**Fig. 33:** Schematic drawing depicting (left) submonolayer cryosorption, where the mechanism is molecule–substrate interaction; and (right) cryocondensation, where the leading mechanism is intermolecular interaction.

– *Cryotrapping.* In this mechanism, the molecules of a low-boiling-temperature gas are trapped in the condensation layer of another gas. This is possible because the interaction energy between dissimilar molecules may be much higher than that between similar molecules. The trapped gas has a saturated vapour pressure several orders of magnitude lower than in its pure condensate. Typical examples are Ar trapped in $CO_2$ at 77 K and $H_2$ in $N_2$ at 20 K.

Modern cryopumps exploit the first two mechanisms. Cryocondensation takes place on a cold surface, in general at 80 K for water vapour and at 10–20 K for the other gas species. The cryosorption of He, $H_2$ and Ne is localized on a hidden surface coated with a porous material. This part of the pump is kept out of the reach of the other type of molecules, i.e. they have a probability close to one to be intercepted and adsorbed by another surface before reaching the cryosorber (see Fig. 34). The cooling is obtained by He gas cooled by a Gifford–McMahon cryocooler.

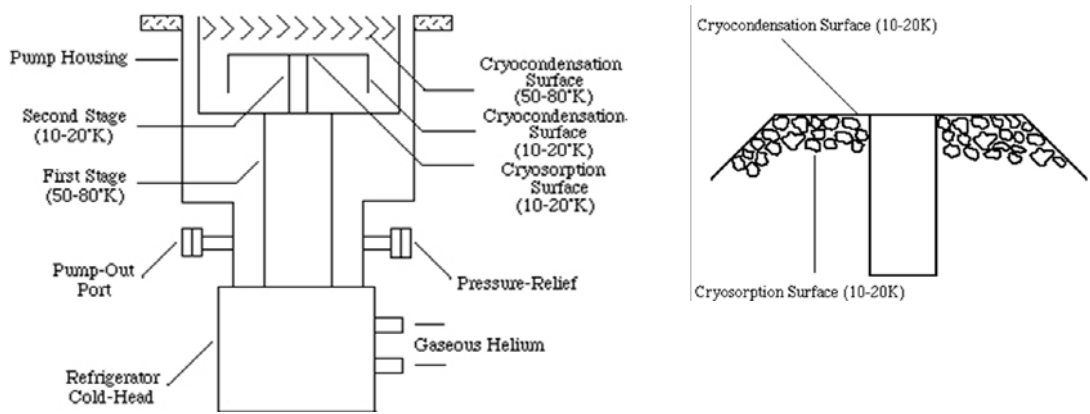

**Fig. 34:** Drawing of a generic cryopump with a closer view to the cryosorption surface where the porous material is fixed.

Cryopumps having pumping speeds in the range 800–60000 l s$^{-1}$ are commercially available (Table 14). For condensable gas molecules, the capture probability is close to 1 (e.g. for water vapour). The maximum gas capacity (also called maximum gas intake) for the condensable gas is limited only by the thermal conductivity of the condensate. To avoid a thick condensate layer and excessive thermal load, cryopumps should be started in the molecular regime (P < 10$^{-3}$ mbar). The quantity and properties of the porous material determine the maximum gas intake of a cryosorbed gas. In general, it is orders of magnitude lower than that for a condensable gas.

**Table 14:** Pumping speeds and maximum gas capacities of a commercial cryopump (Oerlikon-Leybold 800 BL UHV); the pump inlet diameter is 160 mm. Courtesy of Oerlikon-Leybold.

|  | $H_2O$ | $N_2$ | Ar | $H_2$ | He |
| --- | --- | --- | --- | --- | --- |
| $S$ [l s$^{-1}$] | 2600 | 800 | 640 | 1000 | 300 |
| Capacity [Torr l] |  | 225 000 | 225 000 | 3225 | 375 |

Cryopumps require periodic regeneration to evacuate the gas adsorbed or condensed; in this way, the initial pumping speed is fully recovered. To do so, the cryopumps are warmed up to room temperature and the released gas is removed by mechanical pumps (mobile TMP in particle accelerators). During regeneration, the pump is separated from the rest of the system by a valve.

Excessive gas adsorption on the cryosorber leads to performance deterioration. A partial and much faster regeneration (1 h against more than 10 h) may be carried out at temperatures lower than 140°C in such a way as to remove the sorbed gas without releasing water vapour from the pump stage at higher temperature.

## 4.3  Comparison of pumps for ion sources

All pumps presented in the previous sections have been used (or are going to be used) in the vacuum systems of ion sources and in their adjacent vacuum sectors. The choice is based not only on the gas load and required pressure, but also on the costs of the pumps, their maintenance, the duty time of the source and safety aspects. The advantages and disadvantages of the four types of pumps presented in this chapter are summarized in Table 15.

**Table 15:** Advantages and disadvantages of the presented types of pumps.

| Type of pump | Advantages | Disadvantages |
|---|---|---|
| TMP | – No memory effects<br>– Constant pumping speed for pressures lower than $10^{-3}$ mbar<br>– Pumping speed independent of total gas load<br>– Starts working at high pressures (molecular regime) | – Mechanical fragility<br>– Risk of contamination from the backing pump<br>– Need of venting any time the pump is stopped<br>– Need of valve on the inlet flange<br>– Intrinsic limitation in ultimate pressure of $H_2$<br>– Possible vibrations<br>– Regular maintenance |
| SIP | – Clean pumping<br>– No maintenance<br>– No vibrations<br>– Installation in any orientation<br>– Relatively long lifetime<br>– Relatively low cost<br>– Limited but high $H_2$ capacity<br>– The pump current gives a pressure reading | – Low capture probability<br>– Gas selectivity and limited capacity<br>– Memory effects (in particular for rare gases)<br>– Ignition not recommended in the $10^{-3}$–$10^{-4}$ mbar range<br>– Heavy due to permanent magnets<br>– Difficult starting for old pumps<br>– Production of charged particles in particular at start-up<br>– Field emission problems for old pumps<br>– Fringing magnetic field<br>– Safety issue: high voltage |
| Sublimation | – Clean vacuum<br>– High pumping speed for reactive gases<br>– With SIP, extremely low vacuum can be achieved<br>– Low cost per l s$^{-1}$<br>– Electrical power only for sublimation; it works in case of power cut<br>– Limited maintenance (filament change)<br>– No vibration | – Very limited surface gas capacity<br>– Need frequent sublimations at high pressure<br>– Ti film peel-off for high sublimation rates<br>– Selective pumping (no pumping of rare gases and methane)<br>– Risk of leakage current in high-voltage insulators |
| NEG | – Clean vacuum<br>– High pumping speed for reactive gases<br>– With SIP, extremely low vacuum can be achieved<br>– High gas capacity for porous NEG<br>– Low cost per l s$^{-1}$<br>– Electrical power needed only for activation; it works in case of power cut<br>– No maintenance<br>– No vibration | – Selective pumping (no pumping of rare gases and methane)<br>– $H_2$ embrittlement if regeneration cannot be applied<br>– Formation of dust particles is not excluded<br>– Safety issue: pyrophoric, it burns when heated in air at high temperature |
| Cryo | – Very large pumping speed for all gases<br>– Clean vacuum<br>– High pumping capacity<br>– Limited selectivity | – Cost and maintenance<br>– Relatively large volume needed (including refrigerator)<br>– Gas release in case of power cut<br>– Reduced pumping efficiency for $H_2$ for high quantity of gas adsorbed: regeneration needed<br>– Need of valve on the inlet flange |

# 5     The vacuum layout of the Linac4 H⁻ source

Linac4 is a linear accelerator, under construction at CERN, conceived in the framework of the high-intensity and high-luminosity upgrade of the LHC. It will be connected to the LHC injection chain at the latest during the LHC long shutdown 2 (LS2) [25].

In the first sections of Linac4, the main gas loads are the pulsed $H_2$ injection in the source and the continuous injection of the same gas in the Low Energy Beam Transport (LEBT) for beam charge compensation. The vacuum system must evacuate the injected gas and ensure a pressure that is lower than $5 \times 10^{-7}$ mbar in the Radio Frequency Quadrupole (RFQ).

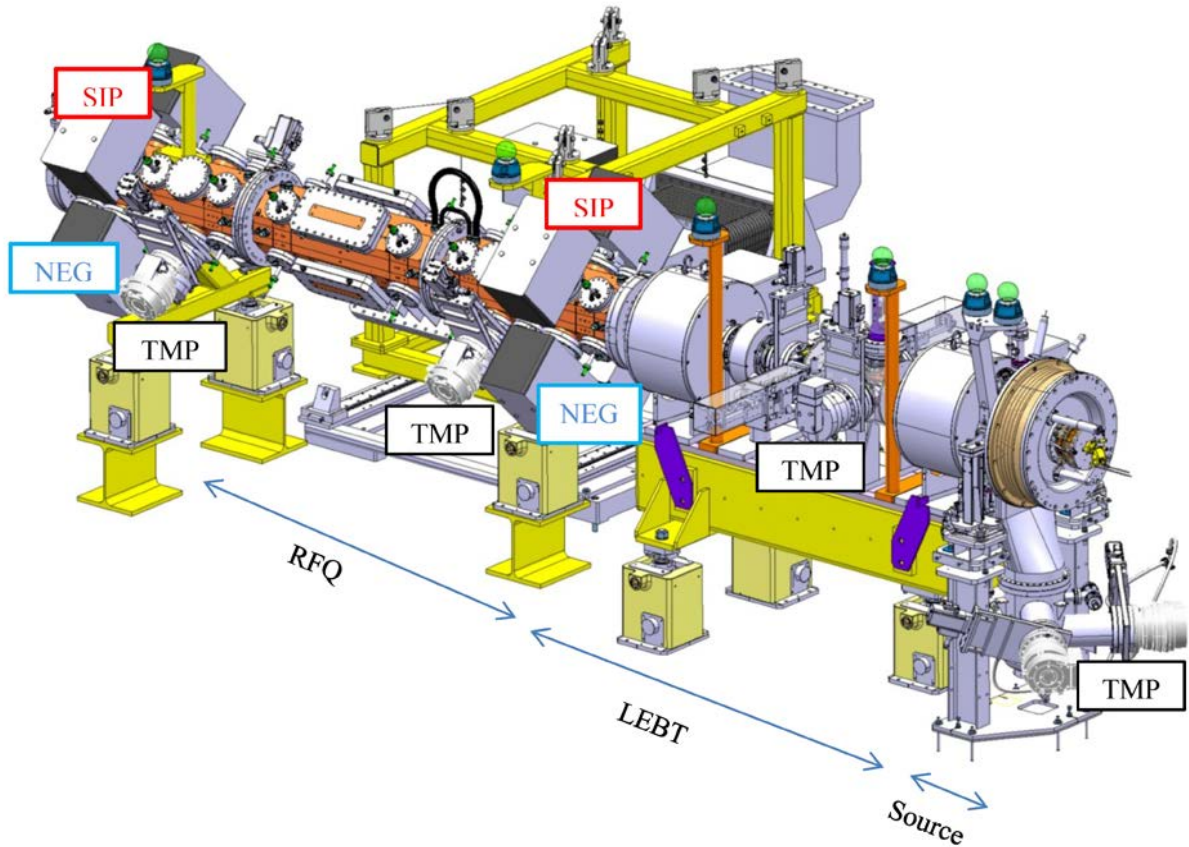

**Fig. 35:** Linac4 layout from the source to the RFQ. Courtesy of D. Steyaert, CERN EN-MME

Figure 35 shows the layout of the source up to the end of the RFQ. The extraction region of the source is equipped with two TMPs that evacuate most of the pulsed flux from the source. An additional TMP is connected to the LEBT diagnostic tank to pump the local continuous injection of H2. The RFQ is divided into three tanks. SIP, NEG pumps and TMP are installed on the first and the third tanks. The NEG pumps provide high pumping speed and capacity for hydrogen. The SIPs contribute to the H2 pumping and ensure the removal of gas species that are not pumped by getters. The TMP are backup pumps in case a problem occurs.

The electrical–network vacuum analogy was applied to evaluate the time-dependent pressure profiles from the source to the end of the RFQ. Figures 36 and 37 show the H2 partial pressure profiles as a function of time and position. Two situations are illustrated, i.e. the pressure profiles with and without continuous injection in the LEBT. A detailed analysis of the vacuum system of the Linac4 source can be found in Ref. [16].

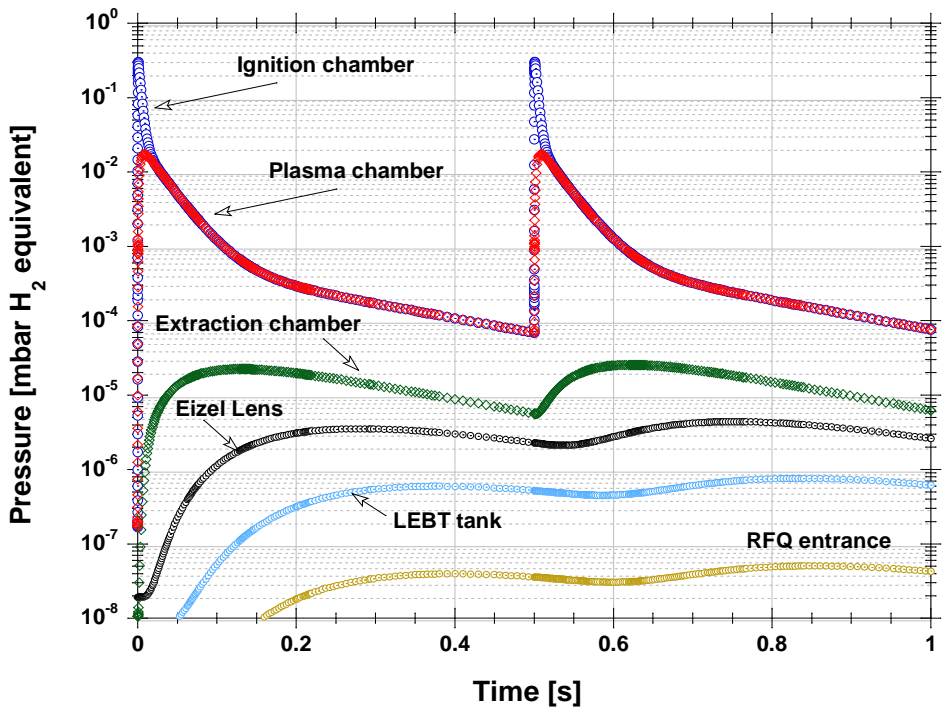

**Fig. 36:** Calculated $H_2$ pressure variations in the Linac4 $H^-$ source. The $H_2$ gas burst from the piezo valve is about $5 \times 10^{-3}$ mbar l per pulse. Each pulse lasts $5 \times 10^{-4}$ s at a repetition frequency of 2 Hz. In this example, the injection in the LEBT is not taken into account.

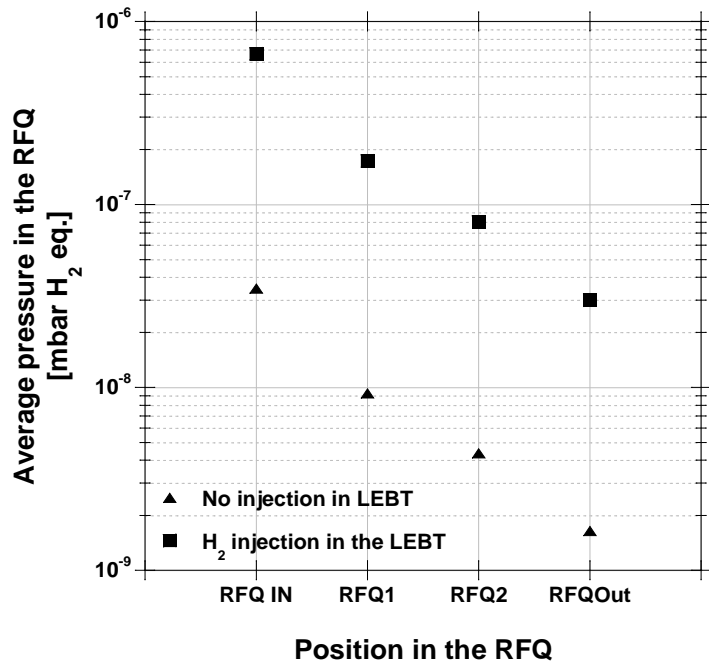

**Fig. 37:** Pressure profiles in the RFQ of Linac4 with and without $H_2$ injection in the LEBT. For the calculation, the injection in the LEBT is equivalent to fixing the pressure in the LEBT to $10^{-5}$ mbar ($H_2$ equivalent).

## Acknowledgement

I wish to thank all my colleagues of the Vacuum, Surfaces and Coatings group of the Technology Department at CERN, including students and visitors. The VSC group has become a centre of excellence in Europe for vacuum technology encompassing all aspects, from surface treatments to computation, operation and control of unique facilities. A significant part of my experience originates from years of work I have been sharing with them. Chiara Pasquino prepared Figs. 36 and 37 and wrote part of section 5.